%% file: paper.tex
\documentclass[11pt]{article}
\usepackage[margin=1.2in]{geometry}

\usepackage{commands}

\begin{document}
\title{Online Team Formation under Different Synergies}

\author{%
  Matthew Eichhorn \\
  Cornell University \\
  Ithaca, NY 14850 \\
  \texttt{meichhorn@cornell.edu} \\
  \and
  Siddhartha Banerjee \\
  Cornell University\\
  Ithaca, NY 14850 \\
  \texttt{sbanerjee@cornell.edu} \\
  \and
  David Kempe \\
  University of Southern California\\
  Los Angeles, CA 90089 \\
  \texttt{david.m.kempe@gmail.com} \\
}

\date{}

\maketitle              
\begin{abstract}%
    Team formation is ubiquitous in many sectors: education, labor markets, sports, etc. A team's success depends on its members' latent types, which are not directly observable but can be (partially) inferred from past performances. From the viewpoint of a principal trying to select teams, this leads to a natural exploration-exploitation trade-off: retain successful teams that are discovered early, or reassign agents to learn more about their types? We study a natural model for online team formation, where a principal repeatedly partitions a group of agents into teams. Agents have binary latent types, each team comprises two members, and a team's performance is a symmetric function of its members' types. Over multiple rounds, the principal selects matchings over agents and incurs regret equal to the deficit in the number of successful teams versus the optimal matching for the given function. Our work provides a complete characterization of the regret landscape for all symmetric functions of two binary inputs. In particular, we develop team-selection policies that, despite being agnostic of model parameters, achieve optimal or near-optimal regret against an adaptive adversary.
\end{abstract}


\import{./}{s1_intro.tex} 
\import{./}{s2_model.tex}
\import{./}{s3_ud.tex}

\import{./}{s4_or.tex}
\import{./}{s5_and.tex}

\section{Conclusion}

Our work provides near-optimal regret guarantees for learning an optimal matching among agents under any symmetric function of two binary variables. While our results are specific to each function, they exhibit several noteworthy common features. First, although we consider an adaptive adversary, it is not hard to see that the regret bounds with i.i.d.~Bernoulli types can only improve by a small constant factor; such a small gap between stochastic and adversarial models is uncommon. Next, for all our settings, minimizing regret turns out to require maximal exploitation (in contrast to quickly learning all agent types, which would benefit from more exploration). Finally, the problems appear to get harder for $\K=\frac{\N}{2}$, and also handling the weakest link setting (i.e., the Boolean $\textsf{AND}$ function) is more challenging than other synergy functions. These phenomena hint at underlying information-theoretic origins, and formalizing these may help in reasoning about more complex models.

Our work raises three natural future directions:
\begin{enumerate}
    \item It would be desirable to close the gaps between our bounds for the \textsf{OR} and \textsf{AND} settings. In each case, however, our results suggest that the optimal procedures may depend heavily on number-theoretic properties of $\N$ and $\K$ which can expose a further level of complication.
    \item We consider only perfect feedback, which in itself presented interesting challenges, but may be unrealistic in real-world settings. Our results likely extend to some noisy feedback models by repeatedly playing a team and averaging their scores. However, quantifying the relationship between the amount of noise and the expected additional regret is an open problem.
    \item The restriction to teams of size 2, and binary agent types, are the main restrictions of our model. For a more general theory of team formation, it is desirable to consider larger teams and other synergy functions (in particular, threshold functions); doing so is a rich and challenging open direction.
\end{enumerate}

\subsubsection*{Acknowledgments}

We would like to thank anonymous reviewers for useful feedback. Part of the work was done when SB and ME were visiting the Simons Institute for the Theory of Computing for the semester on Data-Driven Decision Processes; they also acknowledge support from the NSF under grants ECCS-1847393 and CNS-195599, and the ARO MURI grant W911NF1910217. 
DK acknowledges support from ARO MURI grant ARO W911NF1810208.

%
%
\bibliographystyle{plain}
\bibliography{davids-bibliography/names,davids-bibliography/conferences,davids-bibliography/bibliography,davids-bibliography/publications,refs}

\appendix

\newpage
\import{./}{sa4_ring_details.tex}

\newpage
\import{./}{sa5_ring_proof.tex}
\newpage
\import{./}{sa6_104.tex}

\end{document}

%% file: s1_intro.tex
\section{Introduction} \label{sec:intro}

An instructor teaching a large online course wants to pair up students for assignments. The instructor knows that a team performs well as long as at least one of its members has some past experience with coding, but unfortunately, there is no available information on the students' prior experience. 
However, the course staff can observe the \emph{performance} of each team on assignments, and so, over multiple assignments, would like to reshuffle teams to try and quickly maximize the overall number of successful teams. How well can one do in such a situation?

Team formation is ubiquitous across many domains: homework groups in large courses, workers assigned to projects on online labor platforms, police officers paired up for patrols, athletes assigned to teams, etc. Such teams must often be formed without prior information on each individual's latent skills or personality traits, albeit with knowledge of how these latent traits affect team performance. The lack of information necessitates a natural trade-off: a principal must decide whether to exploit successful teams located early or reassign teammates to gain insight into the abilities of other individuals. The latter choice may temporarily reduce the overall rate of success. 

To study this problem, we consider a setting (described in detail in \cref{sec:model}) where agents have binary latent types, each team comprises two members,  and the performance of each team is given by the same \emph{synergy function}, i.e., some given symmetric function of its members' types. 
Over multiple rounds, the principal selects matchings over agents, with the goal of minimizing the cumulative \emph{regret}, i.e., the difference between the number of successful teams in a round versus the number of successful teams under an optimal matching. 
Our main results concern the special case of symmetric Boolean synergy functions --- in particular, we study the functions \textsf{EQ} and \textsf{XOR} (in \cref{sec:warmup}), \textsf{OR} (in \cref{sec:or}) and \textsf{AND} (in \cref{sec:and}). While this may at first appear to be a limited class of synergy functions, in \cref{sec:robust}, we argue that these four functions are in a sense the \emph{atomic primitives} for this problem; our results for these four settings are sufficient to handle \emph{arbitrary symmetric synergy functions}.

The above model was first introduced by Johari et al.~\cite{Johari18_Team}, who considered the case where agent types are i.i.d.~Bernoulli($p$) (for known $p$) and provide asymptotically optimal regret guarantees under \textsf{AND} (and preliminary results for \textsf{OR}). 
As with any bandit setting, it is natural to ask whether one can go beyond a stochastic model to admit \emph{adversarial} inputs. In particular, the strongest adversary one can consider here is an \emph{adaptive adversary}, which observes the choice of teams in each round, and only then fixes the latent types of agents. In most bandit settings, such an adversary is too strong to get any meaningful guarantees; among other things, adaptivity precludes the use of randomization as an algorithmic tool, and typically results in every policy being as bad as any other. 
Nevertheless, in this work, we provide a \emph{near-complete characterization of the regret landscape for team formation under an adaptive adversary}. In particular, in a setting with $\N$ agents of which $\K$ have type `1', we present algorithms that are agnostic of the parameter $\K$, and yet when faced with an adaptive adversary, achieve optimal regret for \textsf{EQ} and \textsf{XOR}, and near-optimal regret bounds under \textsf{OR} and \textsf{AND} (and therefore, using our reduction in~\cref{sec:robust}, achieve near-optimal regret for any symmetric function).

While our results are specific to particulars of the model, they exhibit several noteworthy features. First, despite the adversary being fully adaptive, our regret bounds differ only by a small constant factor from prior results for \textsf{AND} under i.i.d.~Bernoulli types~\cite{Johari18_Team}; such a small gap between stochastic and adversarial bandit models is uncommon and surprising. Next, our bounds under different synergy functions highlight the critical role of these functions in determining the regret landscape. Additionally, our algorithms expose a sharp contrast between learning and regret minimization in our setting: while the rate of learning increases with more exploration, minimizing regret benefits from maximal exploitation. Finally, to deal with adaptive adversaries in our model, we use techniques from extremal graph theory that are atypical in regret minimization; we hope that these ideas prove useful in other complex bandit settings.

\subsection{Related Work}

Regret minimization in team formation, although reminiscent of \emph{combinatorial bandits/semi-bandits}~\cite{cesa2012combinatorial,Chen2013CombinatorialMB,combes2015combinatorial,gai2012combinatorial,kveton2015tight}, poses fundamentally new challenges arising from different synergy functions.
In particular, a crucial aspect of bandit models is that rewards and/or feedback are linear functions of individual arms' latent types.
Some models allow rewards/feedback to be given by a non-linear \emph{link} function of the sum of arm rewards~\cite{devanur2012online,han2021adversarial}, but typically require the link function to be well-approximated by a linear function~\cite{merlis2020tight}.
In contrast, our team synergy functions are \emph{non-linear}, and moreover, are not well-approximated by any non-linear function of the sums of the agents' types.

One way to go beyond semi-bandit models and incorporate pairwise interactions is by assuming that the resulting reward matrix is low-rank~\cite{katariya2017stochastic,sentenac2021pure,zimmert2018factored}. The critical property here is that under perfect feedback, one can learn all agent types via a few `orthogonal' explorations; this is true in our setting under the \textsf{XOR} function (\cref{sec:warmup}), but not for other Boolean functions.
Another approach for handling complex rewards/feedback is via a Bayesian heuristic such as Thompson sampling or information-directed sampling~\cite{gopalan2014thompson,kirschner2020information,russo2014learning,wang2018thompson}. While such approaches achieve near-optimal regret in many settings, the challenge in our setting is in updating priors over agents' types given team scores. We hope that the new approaches we introduce could, in the future, be combined with low-rank decomposition and sampling approaches to handle more complex scenarios such as shifting types and corrupted feedback.

In addition to the bandit literature, there is a parallel stream on learning for team formation. Rajkumar et al.~\cite{Rajkumar17_Partition} consider the problem of learning to partition workers into teams, where team compatibility depends on individual types. Kleinberg and Raghu~\cite{kleinberg2015team} consider the use of individual scores to estimate team scores and use these to approximately determine the best team from a pool of agents.
Singla et al.~\cite{singla2015learning} present algorithms for learning individual types to form a single team under an online budgeted learning setting.
These works concentrate on pure learning. In contrast, our focus is on minimizing regret. 
Finally, there is a line of work on strategic behavior in teams, studying how to incentivize workers to exert effort~\cite{babaioff2006combinatorial,carlier2010matching}, and how to use signaling to influence team formation~\cite{hssaine2018information}. While our work eschews strategic considerations, it suggests extensions that combine learning by the principal with strategic actions by agents.

%% file: s2_model.tex
\section{Model} \label{sec:model}

\subsection{Agents, Types, and Teams}

We consider $\N$ \emph{agents} who must be paired by a principal into \emph{teams of two} over a number of rounds; throughout, we assume that $\N$ is even.
Each agent has an unknown latent \emph{type} $\type_i\in\{0,1\}$. These types can represent any dichotomous attribute: ``left-brain'' vs.~``right-brain'' (\cref{sec:warmup}), ``low-skill'' vs.~``high-skill'' (\cref{sec:or,sec:and}), etc. We let $\K$ denote the number of agents with type 1, and assume that $\K$ is fixed \textit{a priori} but unknown.

In each round $t$, the principal selects a matching $M_t$, with each edge $(i,j)\in M_t$ representing a \emph{team}. We use the terms ``edge'' and ``team'' interchangeably. The \emph{success} of a team $(i,j)\in M_t$ is $ f(\type_i,\type_j)$, where $f: \{0,1\}^2 \to \mathbb{R}$ is some known symmetric function of the agents' types. In Sections \ref{sec:warmup}-\ref{sec:and}, we restrict our focus to Boolean functions, interpreting $f(\type_i,\type_j) = 1$ as a success and $f(\type_i,\type_j) = 0$ as a failure. The algorithm observes the success of each team, and may use this to select the matchings in subsequent rounds; however, the algorithm cannot directly observe agents' types. For any matching $M$, we define its \emph{score} as $S(M) := \sum_{(i,j)\in M} f(\type_i,\type_j)$ --- in the special case of Boolean functions, this is the number of successful teams. 

A convenient way to view the Boolean setting is as constructing an edge-labeled \emph{exploration graph} $G(V,E_1,$ $E_2,\ldots)$, where nodes in $V$ are agents, and the edge set $E_t := \bigcup_{t' \leq t} M_{t'}$ represents all pairings played up to round $t$. Upon being played for the first time, an edge is assigned a label $\{0,1\}$ corresponding to the success value of its team. \emph{Known 0-agents} and \emph{known 1-agents} are those whose types can be inferred from the edge labels. The remaining agents are \emph{unknown}. The \emph{unresolved subgraph} is the induced subgraph on the unknown agents.

\subsection{Adversarial Types and Regret}
The principal makes decisions facing an \emph{adaptive adversary}, who knows $\K$ (unlike the principal, who only knows $\N$) and, in each round, is free to assign agent types \emph{after seeing the matching chosen by the principal}, as long as (1) the assignment is consistent with prior observations (i.e., with the exploration graph), and (2) the number of 1-agents is $\K$. Note that this is the strongest notion of an adversary we can consider in this setting; in particular, since the adversary is fully adaptive and knows the matching \emph{before} making decisions, randomizing does not help, and so it is without loss of generality to consider only deterministic algorithms.

We evaluate the performance of algorithms in terms of additive \emph{regret} against such an adversary.
Formally, let $M^*$ be any matching maximizing $S(M^*)$ --- note that for any Boolean team success function, $S(M^*)$ is a fixed function of $\N$ and $\K$. 
In round $t$, an algorithm incurs regret $r_t := S(M^*) - S(M_t)$, and its total regret is the sum of its per-round regret over an {a priori infinite time horizon}. Note, however, that after a finite number of rounds, a na\"{i}ve algorithm that enumerates all matchings can determine, and henceforth play, $M^*$; thus, the optimal regret is always finite. Moreover, the ``effective'' horizon (i.e., the time until the algorithm learns $M^*$) of our algorithms is small.

\subsection{Symmetry Synergy Functions and Atomic Primitives} 
\label{sec:robust}

In subsequent sections, we consider the problem of minimizing regret under four Boolean synergy functions $f \colon \{0,1\}^2 \to \{0,1\}$: \textsf{EQ}, \textsf{XOR}, \textsf{OR}, and \textsf{AND}. 
Interestingly, the algorithms for these four settings suffice to handle any symmetric synergy function $f \colon \{0,1\}^2 \to \mathbb{R}$. 
We argue this below for synergy functions that take at most two values; We handle the case of synergy functions $f$ taking three different values at the end of~\cref{ssec:uniform}. 

\begin{lemma}
\label{lem:robust1}
Fix some $\ell \leq u$, let $f \colon \{0,1\}^2 \to \{\ell,u\}$ be any symmetric synergy function,  
and let $r^{f}(n,k)$ denote the optimal regret with $n$ agents, of which $k$ have type $1$. 

Then, $r^{f}(n,k) = (u - \ell)\cdot r^g(n,k)$ for one of $g\in\{\textsf{EQ,XOR,AND,OR}\}$.
\end{lemma}

\begin{proof}
First, note that without loss of generality, we may assume that $f(0,0) \leq f(1,1)$. Otherwise, we can swap the labels of the agent types without altering the problem. Note that this immediately allows us to reduce team formation under the Boolean \textsf{NAND} and \textsf{NOR} function to the same problem under \textsf{AND} and \textsf{OR}, respectively. Next, note that if $f(0,0) = f(1,0) = f(1,1)$, then the problem is trivial, as all matchings have the same score. Otherwise, we may apply the affine transformation $f \mapsto \frac{1}{u-\ell} \cdot f - \frac{\ell}{u-\ell}$ to the output to recover a Boolean function:
\begin{itemize}
    \item When $f(0,1) < f(0,0) = f(1,1)$, we recover the $\textsf{EQ}$ function.
    \item When $f(0,0) = f(1,1) < f(0,1)$, we recover the $\textsf{XOR}$ function.
    \item When $f(0,0) =f(0,1) < f(1,1)$, we recover the $\textsf{AND}$ function.
    \item When $f(0,0) < f(0,1) = f(1,1)$, we recover the $\textsf{OR}$ function.
\end{itemize}
The structure of the problem remains unchanged since total regret is linear in the number of each type of team played over the course of the algorithm. 
The regret simply scales by a factor of $u - \ell$.
\end{proof}

%% file: s3_ud.tex
\section{Uniform and Diverse Teams} \label{sec:warmup}

We first focus on forming teams that promote uniformity (captured by the Boolean \textsf{EQ} function) or diversity (captured by the \textsf{XOR} function). In addition, we also show that the algorithm for $\textsf{EQ}$ minimizes regret under any general symmetric synergy function taking three different values.

\subsection{Uniformity (\textsf{EQ})} \label{ssec:uniform}

We first consider the \emph{equality} (or \textsf{EQ}) synergy function, $f^{\textsf{EQ}}(\type_i,\type_j)=\overline{\type_i\oplus\type_j}$.
Here, an optimal matching $M^*$ includes as few $(0,1)$-teams as possible, and thus
$S(M^*) = \frac{\N}{2} - (\K \!\!\mod 2)$. If $\K$ (and thus $\N-\K$) is even, then all agents can be paired in successful teams; else, any optimal matching must include one unsuccessful team with different types. 
For this setting, \Cref{thm:xnor} shows that a simple policy (\cref{alg:xnor}) achieves optimal regret for \emph{all} parameters $\N$ and~$\K$. 
\begin{algorithm}
\caption{\textsc{Form Uniform Teams}}
\label{alg:xnor}
\begin{algorithmic} 
    \STATE \textbf{Round 1:} Play an arbitrary matching.
    \STATE \textbf{Round 2:} Swap unsuccessful teams in pairs as $\{ (a,b),$ $(c,d) \} \rightarrow \{ (a,c), (b,d) \}$. Repeat remaining teams (including one unsuccessful team when $\K$ is odd).
    \STATE \textbf{Round 3:} If $\{ (a,b), (c,d) \}$, $\{ (a,c), (b,d) \}$ are both unsuccessful, play $\{ (a,d), (b,c) \}$. Repeat remaining teams.
\end{algorithmic}
\end{algorithm}

\begin{theorem} \label{thm:xnor}
    Define $\displaystyle \reg{\textsf{EQ}}{\N}{\K} := 2\cdot\big(\min(\K,\N-\K) - (\K \!\!\mod 2) \big).$ Then,
    \begin{enumerate}[nosep]
        \item \cref{alg:xnor} learns an optimal matching by round 3, and incurs regret at most $\reg{\textsf{EQ}}{\N}{\K}$.
        \item Any algorithm incurs regret at least $\reg{\textsf{EQ}}{\N}{\K}$ in the worst case.
    \end{enumerate}
\end{theorem}

\begin{proof}
  For the upper bound on the regret, note that every unsuccessful team includes a 0-agent and a 1-agent. Thus, there is a re-pairing of any two unsuccessful teams that gives rise to two successful teams. If the re-pairing in round 2 is unsuccessful, the only other re-pairing, selected in round~3, must be successful. There will be $\K \!\!\mod 2$ unsuccessful teams in round 3, making it an optimal matching. At most $\min(\K,\N-\K)$ $(0,1)$-teams can be chosen in each of rounds 1--2, implying that the maximum regret in each of these rounds is  $\min(\K,\N-\K) - (\K \!\!\mod 2)$. Since~\cref{alg:xnor} incurs regret only in rounds 1--2, its total regret is at most $\reg{\textsf{EQ}}{\N}{\K}$. 
    
  For the converse (Claim 2), we argue that against \emph{any} algorithm, the adversary can always induce regret $\min(\K,\N-\K) - (\K \!\!\mod 2)$ in each of rounds 1--2. Note that after round 2, the exploration graph is a union of two (not necessarily disjoint) matchings, and hence consists of a disjoint union of even-length cycles and isolated (duplicated) edges; this is independent of the algorithm, as it holds for any pair of perfect matchings. Since the graph is bipartite, the adversary can assign types such that no pair of the minority type is adjacent in the graph by starting with the labeling according to the bipartition, then arbitrarily relabeling a subset of the minority side to make the labeling consistent with $\K$.
\end{proof}

A similar argument allows us to complete our treatment of general (symmetric) synergy functions from~\cref{sec:robust}.
\begin{corollary}
    For any symmetric synergy function $f:\{0,1\}^2\rightarrow\mathbb{R}$ such that $f(0,0)\neq f(0,1) \neq f(1,1)$, there is a regret-minimizing algorithm that locates an optimal matching within two rounds.
\end{corollary}

\begin{proof}
    By applying an affine transformation to the outputs as in~\cref{sec:robust}, we may assume without loss of generality that $f(0,0)=0$, and $f(1,1)=1$. There are three cases to consider:
    
    \begin{itemize}
        \item $f(0,1) = \frac{1}{2}$: The problem is trivial, since all matchings have the same score.
        \item $f(0,1) > \frac{1}{2}$: The optimal matching includes as many 1-0 agent teams as possible. After the first (arbitrary) matching, every agent is either part of a known 1-0 team or has a known identity (as a member of a 0-0 or 1-1 team). Thus, one can always select an optimal matching in the second round. 
        \item $f(0,1) < \frac{1}{2}$: The optimal matching includes as many 1-1 agent teams as possible, just as in the $\textsf{EQ}$ setting. Note that the three distinct values of $f$ allow us to distinguish between $(0,0)$, $(0,1)$, and $(1,1)$ teams. The same adversarial policy ensures that all 0-1 teams remain sub-optimally paired in round $2$, so we exactly recover the \textsf{EQ} setting.     
    \end{itemize}
\end{proof}

\subsection{Diversity (\textsf{XOR})}

Next we consider the \textsf{XOR} success function, $f^{\textsf{XOR}}(\type_i,\type_j)=\type_i\oplus\type_j$, which promotes diverse teams.
Now $S(M^*) = \min(\K,\N-\K)$, since any optimal matching $M^*$ includes as many $(0,1)$-teams as possible. 
Define $x^+ := \max(0, x)$; we again show that a simple policy (\cref{alg:xor}) has optimal regret for all $\N,\K$.

\begin{algorithm}
\caption{\textsc{Form Diverse Teams}}
\label{alg:xor}
\begin{algorithmic} 
    \STATE \textbf{Round 1:} Play an arbitrary matching;
    let $\{(\agent{1},\agent{2}),\hdots,(\agent{\ell-1},\agent{\ell})\}$ denote  unsuccessful teams.
    \STATE \textbf{Round 2:} Replay all successful teams, and construct a single cycle over all unsuccessful teams (i.e., play teams $\{ (\agent{\ell}, \agent{1}), (\agent{2}, \agent{3}), \hdots, (\agent{\ell-2},\agent{\ell-1}) \}$).
    \STATE \textbf{Round 3:} Play any inferred optimal matching (see \Cref{thm:xor}).
  \end{algorithmic}
\end{algorithm}

\begin{theorem} \label{thm:xor}
    Define $\displaystyle \reg{\textsf{XOR}}{\N}{\K} := 2 \cdot \big( \min(\K,\N-\K) - 1 - (\K \!\!\mod 2) \big)^+.$ Then,
    \begin{enumerate}[nosep]
        \item \cref{alg:xor} learns an optimal matching after round 2, and incurs regret at most $\reg{\textsf{XOR}}{\N}{\K}$.
        \item Any algorithm incurs regret at least $\reg{\textsf{XOR}}{\N}{\K}$ in the worst case.
    \end{enumerate}
\end{theorem}

\begin{proof}
   For the achievability in Claim 1, note that each edge $\big(i, (i+1) \! \mod \ell \big)$ of the cycle constructed in the algorithm has the following property: if the edge is successful in round 2, then its endpoints have opposite types; otherwise, they have the same type. By following edges around the cycle, the algorithm can therefore construct the sets $S^=$ of agents with the same type as agent $\agent{1}$, and $S^{\neq}$ of agents with the opposite type. Subsequently, it is optimal to match $\min(|S^=|, |S^{\neq}|)$ teams of (known) opposite-type agents, and match the extraneous agents into unsuccessful teams.

    Among agents $\agent{1}, \ldots,\agent{\ell}$, there are $\K-\frac{\N-\ell}{2}$ 1-agents and $\N-\K-\frac{\N-\ell}{2}$ 0-agents; thus, the round 1 regret is
    $
        r_1 := \min(\K,\N-\K) - \tfrac{\N-\ell}{2}
    $ (note that When $\K$ is odd, one team must be successful in round 1).
    Since no regret is incurred after round 2, the adversary must maximize the regret in round 2 conditioned on the choice of $\ell$. This is achieved by assigning type 1 to agents $1, \ldots, \K-\frac{\N-\ell}{2}$, and type 0 to agents $\K-\frac{\N-\ell}{2}+1, \ldots, \ell$.
    Since $(\ell,1)$ and $(\K-\frac{\N-\ell}{2}, \K-\frac{\N-\ell}{2}+1)$ are the only successful teams (as long as agents $\agent{1}$ to $\agent{\ell}$ do not all have the same type), the regret in round 2 is $(r_1 - 2)^+$. The total regret $\big(2\min(\K,\N-\K) - \N + \ell - 2\big)^+$ is monotone increasing in $\ell$, with the maximum attained at $\ell = \N - 2 (\K \!\!\mod 2)$. Substituting, we get the upper bound.
    
    For the converse (Claim 2), we describe a policy for the adversary that ensures regret at least $\reg{\textsf{XOR}}{\N}{\K}$. In round 1, the adversary reveals $\K \!\!\mod 2$ successful teams, resulting in regret $\min(\K,\N-\K) - (\K \!\!\mod 2)$. In round 2, the exploration graph must consist of a disjoint union of even-length cycles (including isolated duplicated edges).

    First, when $\K$ is odd, consider the component containing the one revealed successful team from round 1. If the component has just two agents (i.e., the algorithm repeats the team), then we again get one successful team. Otherwise, if the team is part of a longer cycle, the adversary puts an odd number of adjacent 0s and an odd number of adjacent 1s in the cycle, such that the previously successful team is (0,1). Since the edge is not repeated, and only one other (0,1)-team is created, the algorithm gets at most one successful team in this cycle. The remaining cycles contain an even number of 1-agents, so we appeal to below.

    When $\K$ is even, the adversary fills cycles with 0-agents until they are exhausted, then labels all remaining agents as 1-agents. At most one cycle contains both agent types. Placing the 0-agents contiguously in this cycle ensures only two adjacent successful teams. Since all cycle lengths are even, as is $\N-\K$, these successful teams will be an even number of edges apart; in particular, the adversary can ensure that they are both edges from round 2, making the assignment consistent with round 1. In total, the algorithm obtains at most $2 + (\K \!\!\mod 2)$ successful teams in round 2, giving total regret at least $\reg{\textsf{XOR}}{\N}{\K}$.
\end{proof}

%% file: s4_or.tex
\section{The Strongest Link Setting (\textsf{OR})} \label{sec:or}

We next consider the Boolean \textsf{OR} 
synergy function, that is, $f^{\textsf{OR}}(\type_i,\type_j)=\type_i+\type_j$.
Adopting the terminology of Johari et al. \cite{Johari18_Team}, we refer to this setting as the \emph{strongest link} model: interpreting 0/1-agents as having low/high skill, a team is successful when it has at least one high-skill member.

Observe that under \textsf{OR}, we have $S(M^*) = \min(\K,\N/2)$, since any optimal matching $M^*$ includes a maximal set of $(0,1)$-teams. Define $\al := \frac{\N-\K}{\N}$ to be the \emph{fraction of low-skill agents}; our regret bounds in this setting are more conveniently phrased in terms of $\al$.
In particular, our first result establishes the following lower bounds on the regret incurred by \emph{any} algorithm. 

\begin{theorem} \label{thm:orlb}
For the strongest link setting, any algorithm incurs regret at least $L^{\textsf{OR}}(\al) \cdot \N$ in the  worst-case, where 
    \[
        L^{\textsf{OR}}(\al)  = \begin{cases}
            \frac{13\al}{17} & 0 \leq \al \leq \frac{1}{2} \\
            \frac{6-9\al}{4} & \frac{1}{2} < \al \leq \frac{6}{11} \\
            \frac{3-4\al}{3} & \frac{6}{11} < \al \leq \frac{3}{5} \\
            \frac{1-\al}{2} & \frac{3}{5} < \al \leq 1
        \end{cases}
    \]
\end{theorem}


We establish the bound given in this theorem via a sequence of lemmas that present and analyze an adversarial policy for a particular range of $\al$. Since the adversary knows $\al$, he can choose the worst policy, i.e., achieve the pointwise maximum of the regret of the different policies. We begin with a lemma that allows us to restrict our attention to a particular subfamily of algorithms.

\begin{lemma} \label{lem:algdoesntrepeat}
    There exists an optimal algorithm that never pairs known 0-agents until the types of all agents are known.
\end{lemma}

\begin{proof}
    We argue the claim with an exchange argument. Suppose that the algorithm pairs two known 0-agents in a round in which it also plays a team $(u,v)$, where at least one of $u,v$ is unknown. If $u,v$ are both 0-agents, then the algorithm will learn this whether it pairs $u$ and $v$ or explores $u$ and $v$ with the known 0-agents. Both of these actions result in the same regret. If exactly one of $u,v$ is a 0-agent, then the algorithm again accrues no additional regret by exploring $u,v$ as opposed to pairing them. Moreover, the algorithm can only gain more information about the types of $u,v$ through exploration. Finally, if both $u$ and $v$ are 1-agents, then the algorithm incurs one less unit of regret and learns the types of both $u$ and $v$ by exploring them with 0-agents as opposed to pairing them. By repeatedly applying these swaps, we never increase the regret of the algorithm, nor reduce the set of deductions that it can make about node types. When the swapping process finishes, we are left with an algorithm of the claimed type.
\end{proof}

By this lemma, we can, without loss of generality, restrict our attention to algorithms that use all 0-agents for exploration. 
  
\begin{lemma} \label{lem:orlb1}
    There is a policy for the adversary that ensures regret at least $\frac{13\al}{17} \cdot \N$ for all $\al \leq \frac{1}{2}$. 
\end{lemma}

\begin{proof}
  Recall that for $\al \leq \frac{1}{2}$, per-round regret is equal to the number of $(0,0)$-teams that were selected.
  Consider the following adversarial policy:
    
    \noindent\textbf{Round 1:} The algorithm chooses an arbitrary matching. The adversary reveals $\frac{4\al}{17} \cdot \N$ $(0,0)$-teams, giving regret $\frac{4\al}{17} \cdot \N$. 
    
    \noindent\textbf{Round 2:} The algorithm explores with its $\frac{8\al}{17} \cdot \N$ known 0-agents. It will explore members from $\left( \frac{4\al}{17} + \beta \right) \cdot \N$ teams from round 1, for some $0 \leq \beta \leq \frac{4\al}{17}$. The adversary can label one member from each of these teams as a 0-agent, giving regret $\left(\frac{4\al}{17} + \beta \right) \cdot \N$. 
    
    \noindent\textbf{Round 3--4:} To this point, the algorithm has learned the types of $\left( \frac{12\al}{17} + \beta \right) \cdot \N$ 0-agents and $\left( \frac{4\al}{17} + \beta \right) \cdot \N$ 1-agents (the partners of 0-agents from successful teams). Therefore, the algorithm again explores with $\frac{8\al}{17} \N$ known 0-agents in round 3; let $S$ be the set of unknown agents that is being explored, of size $|S| = \frac{8\al}{17} \N$. Let $I \subseteq S$ be a maximum independent set of $S$ in the unexplored subgraph. Since the unexplored subgraph is bipartite, $|I| \geq \frac{|S|}{2} = \frac{4\al}{17} \cdot \N$.

      There remain $\left( \frac{5\al}{17} - \beta \right) \cdot \N$ unknown 0-agents. If $|I| \geq \left( \frac{5\al}{17} - \beta \right) \cdot \N$, then the adversary reveals an arbitrary subset of $I$ of size $\left( \frac{5\al}{17} - \beta \right) \cdot \N$ as 0-agents. This gives regret $\left( \frac{5\al}{17} - \beta \right) \cdot \N$ in round 3, and brings the total accumulated regret to $\frac{13\al}{17} \N$. 
    
    Otherwise, the adversary reveals all of $I$ as 0-agents. We write $|I| = \left( \frac{4\al}{17} + \gamma \right) \cdot \N$, for some $0 \leq \gamma \leq \left( \frac{\al}{17} - \beta \right)$. The regret in round 3 is then $\left( \frac{4\al}{17} + \gamma \right) \cdot \N$.
    
    Since the exploration graph after round 2 is 2-regular, each 0-agent revealed in round 3 can result in the revelation of at most two 1-agents (its neighbors in the exploration graph). Therefore, after round 3, there are still at least $\left( \frac{4\al}{17} - \gamma \right) \cdot \N$ more known 0-agents than known 1-agents. The algorithm uses these excess 0-agents for exploration in round 4. Since the exploration graph is now 3-regular, the adversary can use a greedy construction to select an independent set of at least one fourth of the explored nodes, containing at least $\left( \frac{\al}{17} - \frac{\gamma}{4} \right) \cdot \N$ nodes. The number of remaining unknown 0-nodes at this point is $\left( \frac{\al}{17} - \beta - \gamma \right) \cdot \N$, so the adversary can reveal
      $\min \left( \left( \frac{\al}{17} - \beta - \gamma \right) \cdot \N, \left( \frac{\al}{17} - \frac{\gamma}{4} \right) \cdot \N \right)
      = \left( \frac{\al}{17} - \beta - \gamma \right) \cdot \N$ 0-agents.
This gives regret $\left( \frac{\al}{17} - \beta - \gamma \right) \cdot \N$ in round 4, bringing the total accumulated regret to $\frac{13\al}{17} \N$.
\end{proof}

The remaining lemmas handle the case $\alpha \geq \frac{1}{2}$. Recall that here, the per-round regret is equal to the number of $(1,1)$-teams that were selected. 

\begin{lemma} \label{lem:orlb2}
    The adversary has a strategy that ensures regret at least $\frac{\K}{2}$ for each $\frac{1}{2} \leq \al \leq 1$. 
\end{lemma}

\begin{proof}
    The adversary reveals $\frac{\K}{2}$ $(1,1)$-teams in round 1, giving regret $\frac{\K}{2}$.
\end{proof}

Note that by definition, $\frac{\K}{2} = \frac{1-\al}{2} \cdot \N$. Next, we show that the adversary has an alternate 2-round strategy which establishes a different linear lower bound on the regret.

\begin{lemma} \label{lem:orlb3}
    The adversary has a strategy that ensures regret at least $\frac{3-4\al}{3} \cdot \N$ for each $\frac{1}{2} \leq \al \leq 1$. 
\end{lemma}

\begin{proof}
    In round 1, the adversary reveals $\frac{\al}{3} \cdot \N$ $(0,0)$-teams. The remaining teams are $\frac{\al}{3} \cdot \N$ $(0,1)$-teams and $\frac{3-4\al}{6} \cdot \N$ $(1,1)$-teams, giving regret $\frac{3-4\al}{6} \cdot \N$.
    In round 2, the algorithm uses the $\frac{2\al}{3}$ known 0-agents to explore. It must explore members of at least $\frac{\al}{3}$ teams from round 1. The adversary can label one member each from $\frac{\al}{3}$ of these teams as a 0-agent, giving additional regret $\frac{3-4\al}{6} \cdot \N$ in round 2, and making the total accumulated regret $\frac{3-4\al}{3} \cdot \N$.
\end{proof}

The bound from Lemma~\ref{lem:orlb3} is strictly better than the one from Lemma~\ref{lem:orlb2} for $\frac{1}{2} \leq \al < \frac{3}{5}$.
Finally, we present a 3-round adversary strategy that establishes a third linear regret bound. 

\begin{lemma} \label{lem:orlb4}
    The adversary has a strategy that ensures regret at least $\frac{6-9\al}{4} \cdot \N$ for each $\frac{1}{2} \leq \al \leq 1$. 
\end{lemma}

\begin{proof}
    In round 1, the adversary reveals $\frac{\al}{4} \cdot \N$ $(0,0)$-teams. The remaining teams are $\frac{\al}{2} \cdot \N$ $(0,1)$-teams and $\frac{2-3\al}{4} \cdot \N$ $(1,1)$-teams, giving regret $\frac{2-3\al}{4} \cdot \N$.
    In round 2, the algorithm uses the $\frac{\al}{2} \cdot \N$ known 0-agents to explore, exploring members from $\left( \frac{\al}{4} + \beta \right) \cdot \N$ successful teams from round 1, for some $0 \leq \beta \leq \frac{\al}{4}$. The adversary can label one member from each of these teams as a 0-agent. Thus, the algorithm observes $\left( \frac{\al}{4} + \beta \right) \cdot \N$ (0,0)-teams, $\left(\frac{\al}{2} - 2\beta \right) \cdot \N$ (0,1)-teams, and $\left(\frac{2-3\al}{4} - \beta \right) \cdot \N$ (1,1)-teams, giving the algorithm regret $\left(\frac{2-3\al}{4} - \beta \right) \cdot \N$.
    
    Note that every explored 0-agent in round 2 had been paired with a 1-agent in round 1. Therefore, the algorithm again has $\frac{\al}{2} \cdot \N$ 0-agents with which to explore in round 3, and there are $\left( \frac{\al}{4} - \beta \right)$ unknown 0-agents. Note that the unresolved subgraph is bipartite after round 2, and at least half of the nodes explored by the algorithm must fall on the same side of this bipartition. Therefore, the adversary can reveal at least $\min \left( \frac{1}{2} \cdot \frac{\al}{2} \cdot \N, \left(\frac{\al}{4} - \beta\right) \cdot \N \right) = \left( \frac{\al}{4} - \beta \right) \cdot \N$ 0-agents in round 3. This gives regret $\left(\frac{2-3\al}{4} + \beta \right) \cdot \N$ in round 3, so the total regret is $\frac{6-9\al}{4} \N$.
\end{proof}

The bound from Lemma~\ref{lem:orlb4} is strictly better than those from Lemmas~\ref{lem:orlb2} and \ref{lem:orlb3} for $\frac{1}{2} \leq \al < \frac{6}{11}$. Taking the pointwise maximum of the linear functions in the lemmas, we get~\Cref{thm:orlb}.

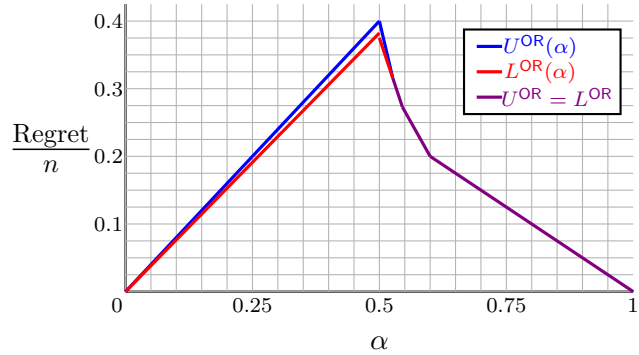
\begin{figure}[t] 
    \centering
    \import{./}{orplot.tex}
    \caption{\small\em Our regret bounds (\Cref{thm:orlb,thm:ub_smallalpha,thm:pwlinearub_improved}) under the Strongest Link model, as functions of $\al := \frac{\N-\K}{\N}$, the fraction of low-skill agents. The bounds match for $\frac{10}{19} \leq \al \leq 1$.}
    \label{fig:orboundsplot}
\end{figure}

The lower bound in \cref{thm:orlb} is plotted in~\cref{fig:orboundsplot}, and notably varies greatly with $\al$. Nevertheless, we provide a policy (\cref{alg:or4clique}) that manages to achieve \emph{nearly matching regret across all $\al$}, while being agnostic of $\K$ (and thus $\al$). Both bounds are plotted in~\cref{fig:orboundsplot}; despite the functions being piecewise linear, they match exactly for $\al \geq \frac{10}{19}$, and $U^{\textsf{OR}}(\al) - L^{\textsf{OR}}(\al) < 0.018$ for all $\al$.

\subsection{The \maxexploit\ Algorithm} \label{sec:or-algorithm}

To simplify our analysis, we introduce some terminology: we say that two unknown 0-agents become \emph{discovered} when they are paired to form an unsuccessful team. An unknown agent is \emph{explored} when its type is revealed by pairing it with a known 0-agent. Our policy for this setting, \maxexploit, is given in \cref{alg:or4clique}. The algorithm exploits the inferred types of agents to the greatest possible extent; a maximal number of known 1-0 agent teams are played in each round. Exploration is only done using known 0-agents that cannot be included in such a pair.
If only two agents in a 4-cycle are explored, we treat the other two agents as unknown, even if their type is deducible.

\begin{algorithm}[!h]
\caption{\maxexploit}
\label{alg:or4clique}
\begin{algorithmic} 
    \STATE \textbf{Round 1}: Select an arbitrary matching.
    \WHILE { $\#\{\text{known 0-agents}\} > \#\{\text{known 1-agents}\}$ \textbf{ and } $\#\{\text{unknown agents}\} > 0$ }
        \STATE Pair each known 1-agent with a known 0-agent.
        \STATE Use extra known 0-agents to explore both members of successful teams. 
        
        (In round 3, explore all members of 4-cycles whenever possible\footnotemark.)
        \STATE \textbf{Round 2:} Re-pair remaining unknown successful teams into 4-cycles. 
        
        (If number of remaining unknown successful teams is odd, repeat one team.)
        \STATE \textbf{Round 3:} In each 4-cycle with undiscovered agents, re-pair to form a 4-clique.
        \STATE \textbf{Round 4+:} Re-play the matching from round 1 on unexplored successful teams.
    \ENDWHILE
\end{algorithmic}
\end{algorithm}

First, to see that the algorithm terminates, note that in each iteration of the loop, at least one known 0-agent is used for exploration, revealing the type of another agent. Thus, the algorithm makes progress and eventually terminates. 

Next, note that unknown agents are always in successful teams throughout the algorithm (as both members of an unsuccessful team can be deduced as 0-agents). Upon termination, the algorithm can play an optimal matching: either all agents are known, or there are enough known 1-agents to match all known 0-agents, and the other successful teams of unknown agents can be safely replayed.
    
Let $d_t$ be the number of 0-agents discovered in round $t$ by pairing two unknown agents,
and $e_t$ the number of 0-agents revealed by exploration with a known 0-agent.
We define
\[
\Diff{t} := \; \#\{\text{known 0-agents after round $t$}\} \; - \#\{\text{known 1-agents after round $t$}\}
\]
The following lemma studies how $\Diff{t}$ evolves over rounds $t$.

\begin{lemma} \label{lem:delta-properties}
$\Diff{1} = d_1$, and $2e_t = \Diff{t}\leq \Diff{t-1}$, for all $t \geq 2$.
\end{lemma}

\begin{proof}
In round 1, the algorithm discovers $d_1$ 0-agents, and no 1-agent (since there is no exploration); hence $\Diff{1} = d_1$. 
Consider the 4-cycle and 4-clique edges played in rounds 2--3. If such an edge comprises two 0-agents, then the other two agents in its cycle or clique must be 1-agents. In particular, the addition of $d_t$ known 0-agents in these rounds is exactly counterbalanced by the deduction of their neighboring $d_t$ 1-agents, so discovery does not contribute to $\Diff{t+1} - \Diff{t}$.

Next, consider any round $t \geq 2$. The algorithm first pairs all known 1-agents with known 0-agents, so exactly $\Diff{t-1}$ agents are used for exploration. Each exploration must discover either a 0-agent or a 1-agent, so
$\Diff{t} = \Diff{t-1} + e_t - (\Diff{t-1} - e_t) = 2e_t$. 
Since members of successful teams are explored in pairs, at most half of all explorations can reveal 0-agents. Thus, $e_t \leq \frac{\Diff{t-1}}{2}$.
\end{proof}

For the subsequent analysis, there are two distinct regimes depending on the \emph{fraction} of low-skill agents $\al$. When most agents are low-skill ($\al > \frac{1}{2}$), the optimal configuration includes some $(0,0)$-teams, but no $(1,1)$-teams, and $r_t$ equals the number of $(1,1)$-teams in $M_t$.
On the other hand, when most agents are high-skill ($\al \leq \frac{1}{2})$, the optimal configuration consists entirely of successful teams, and an algorithm's round-$t$ regret $r_t$ is the number of $(0,0)$-teams in $M_t$.
Consequently, the analysis in each regime is very different.

\subsection[Majority High-Skill Regime]{Majority High-Skill Regime ($\al \leq \frac{1}{2}$)}

We begin the analysis by focusing on the case when $\al \leq \frac{1}{2}$.
Recall that the total regret in this regime equals the total number of $(0,0)$ teams the algorithm plays.

\begin{theorem} \label{thm:ub_smallalpha}
    For $\alpha \leq \tfrac{1}{2}$, \cref{alg:or4clique} has regret at most $\frac{4}{5} \cdot \al\N$.
\end{theorem}

\begin{proof}
    First, note that in the regime $\alpha \leq \frac{1}{2}$, the algorithm never pairs two \emph{known} 0-agents; known 0-agents are paired with known 1-agents or used for exploration. Hence, the number of $(0,0)$-teams selected, and thus the regret, in round $t$ is $e_t + \frac{d_t}{2}$. (Note that $e_1 = 0$.)
    
    After round 3, by Lemma~\ref{lem:delta-properties}, there are $2e_3$ more known 0-agents than 1-agents.
    The unresolved agents are contained in 4-cliques of successful teams, which must each contain at least three 1-agents. Thus, exploring any 0-agent means that the algorithm can deduce three 1-agents. After $e_3$ such explorations, the algorithm locates $3e_3$ 1-agents, terminating the loop. The regret incurred in rounds 4 and later is thus at most $e_3$, giving total regret at most $\frac{d_1}{2} + \frac{d_2}{2} + \frac{d_3}{2} + e_2 + 2e_3$. 
         
    We can now bound the regret incurred by \cref{alg:or4clique} by formulating the adversary's problem of choosing the worst-case number of revealed zeros in each round as an LP with variables $\{d_1,d_2,d_3,e_2,e_3\}$.
    Applying Lemma~\ref{lem:delta-properties} to rounds 2 and 3, we obtain that $e_2 \leq \frac{d_1}{2}$ and $e_3 \leq e_2$.
    In addition, $d_1 + d_2 + d_3 + e_2 + 2e_3 \leq \al\N$ ensures that the number of 0-agents revealed by the adversary is at most the total number of 0-agents. Put together, we get the following LP:
    \begin{align*}
        \textrm{Maximize:} \hspace{20pt} & \tfrac{d_1}{2} + \tfrac{d_2}{2} + \tfrac{d_3}{2} + e_2 + 2e_3 \\
        \textrm{Subject to:} \hspace{20pt} 
        & e_2 \leq \tfrac{d_1}{2} \\
        & e_3 \leq e_2 \\
        & d_1 + d_2 + d_3 + e_2 + 2e_3 \leq \al\N \\
        & d_1, d_2, d_3, e_2, e_3 \geq 0
    \end{align*}
            
    \noindent Solving, we get
    $(d_1, d_2, d_3, e_2, e_3) = (\frac{2\al\N}{5},0,0,\frac{\al\N}{5},\frac{\al\N}{5})$ as the adversary's best strategy, with regret at most $\frac{4}{5}\al\N$.
\end{proof}

\subsection[Majority Low-Skill Regime]{Majority Low-Skill Regime ($\al > \frac{1}{2}$)}

A different, more involved, analysis shows that~\cref{alg:or4clique} is also near-optimal when $\alpha > \frac{1}{2}$. 

\begin{theorem} 
\label{thm:pwlinearub_improved}
For $\al > \frac{1}{2}$, \cref{alg:or4clique} learns an optimal matching after incurring regret at most $U^{\textsf{OR}}(\al) \cdot \N$, where
    \[
        U^{\textsf{OR}}(\al) = \begin{cases}
            \tfrac{10-16\al}{5} & \tfrac{1}{2} \leq \al < \tfrac{10}{19}, \\
            \frac{6-9\al}{4} & \tfrac{10}{19} \leq \al < \tfrac{6}{11}, \\
            \frac{3-4\al}{3} & \tfrac{6}{11} \leq \al < \tfrac{3}{5}, \\
            \frac{1-\al}{2} & \tfrac{3}{5} \leq \al \leq 1.
        \end{cases}
    \]
\end{theorem}

\noindent Note that $\lim_{\al \downarrow \frac{1}{2}} U^{\textsf{OR}}(\al) = \frac{2}{5}$, which matches $\lim_{\al \uparrow \frac{1}{2}} U^{\textsf{OR}}(\al)$ from \Cref{thm:ub_smallalpha}. 
Before proceeding, we define $\zz{t}, \zo{t}, \oo{t}$ to be the number of $(0,0), (0,1)$, and $(1,1)$-teams the algorithm plays in round $t$, respectively. Since there are $(1-\al) \N$ 1-agents in total, $\zo{t} = (1-\al) \N - 2 \oo{t}$; in turn, since there are $\al \N$ 0-agents, $\zz{t} = \frac{1}{2} \cdot (\al \N - \zo{t}) = \oo{t} + (\al - \frac{1}{2}) \cdot \N > \oo{t}$.
We now prove \Cref{thm:pwlinearub_improved} via a series of lemmas. 
 
\begin{lemma} \label{lem:wlogadversary}
The adversary has a best response to~\cref{alg:or4clique} with the following properties:
\begin{enumerate}
    \item It never reveals pairs of unknown agents as $(0,0)$-teams after round 1.
    \item It never reveals any $(1,1)$-team until all $(0,1)$-teams have been revealed.
\end{enumerate}
\end{lemma}

\begin{proof}
    For the first claim, suppose that the adversary reveals a $(0,0)$-team among the re-paired teams in round $t=2$ or $t=3$. The 4-cycle or 4-clique containing this $(0,0)$-team contains two 0-agents and two 1-agents, so two $(0,1)$-teams were selected in each of the first $t-1$ rounds.
    The adversary can force the same regret, and provide the same information, by relabeling these agents so a $(0,0)$-team and a $(1,1)$-team are revealed in round 1, the two 1-agents are explored in round 2, and (if $t=3$) the $(0,1)$-teams are repeated in round 3.
    By repeating this relabeling, we arrive at an adversary strategy of the same regret, of the claimed form.
    
    For the second claim, recall that the algorithm's regret in the regime $\al > \frac{1}{2}$ is exactly the number of $(1,1)$-teams it plays. We will describe a scheme charging $(1,1)$-teams played in round $t$ to $0$-agents explored in round $t$. 
  
    Consider some round $t \geq 2$ in which $\oo{t}$ $(1,1)$-teams are played. Since $\zz{t} > \oo{t}$, and all $(0,0)$-teams result from exploration\footnote{except in the last round, where known (0,0)-teams may be played; however, no (1,1)-teams are played in this round.} by the first claim, we can charge one distinct explored 0-agent for each such $(1,1)$-team. Thus, the number of \emph{uncharged} explored 0-agents in round $t$ is exactly $\zz{t} - \oo{t} = (\al - \frac{1}{2}) \cdot \N$, independent of $t$ and $\oo{t}$. 
  
    The total number of 0-agents explored in rounds $t \geq 2$ is exactly $\zo{1}$.
    If the algorithm runs for $T$ rounds, exactly $(T-1) \cdot (\al - \frac{1}{2}) \cdot \N$ explored 0-agents remain uncharged. Thus, the number of \emph{charged} 0-agents, which equals the regret incurred after round 1, is $\zo{1} - (T-1) \cdot (\al - \frac{1}{2}) \cdot \N$.
    Conditioned on $\zz{1}, \zo{1}, \oo{1}$, the regret is therefore maximized by minimizing $T$; that is, the adversary wants the algorithm to finish in as few rounds as possible.
    To minimize the number of rounds $T$, the adversary should maximize the number of 0-agents available for exploration. The adversary accomplishes this by having the algorithm explore $(0,1)$-teams before any $(1,1)$-teams.
\end{proof}

\noindent We now focus, without loss of generality, on such an adversary. This lets us bound the regret in terms of $(\zz{1}, \zo{1}, \oo{1})$.
  
\begin{lemma} \label{lem:regretbound}
    Conditioned on $\zz{1}, \zo{1}, \oo{1}$, \cref{alg:or4clique} has regret at most 
    \[
        \floor{\frac{\zo{1} + \oo{1}}{\zz{1}}} \oo{1} + \min(\oo{1}, (\zo{1} + \oo{1}) \! \mod \zz{1}).
    \]
\end{lemma}

\begin{proof}
From Lemma~\ref{lem:wlogadversary} we know that only $(0,1)$-teams are explored before any $(1,1)$-team is explored.
Each explored $(0,1)$-team results in pairing a 0-agent with a newly discovered 1-agent, and also adds a known 0-agent. Thus as long as the algorithm explores only $(0,1)$-teams, the number of pairs of 0-agents available for exploration stays constant at $\zz{1}$.
Therefore, the total number of rounds of exploration until all agents in successful teams are explored is $\ceil{\frac{\zo{1} + \oo{1}}{\zz{1}}}$. One subtlety here is that the exploration of $(1,1)$-teams decreases the available 0-agents. However, because $\oo{1} < \zz{1}$, the first round in which a $(1,1)$-team can be explored is one round before the last; in this case, the last round only explores $(1,1)$-teams, and the number of 0-agents available for this exploration exceeds the number of 1-agents to be explored. Thus, the bound on the number of rounds of exploration does hold.
In the last round of exploration, no $(1,1)$-teams can be played, so no regret is incurred.
We therefore focus on the first $T = \floor{\frac{\zo{1} + \oo{1}}{\zz{1}}}$ rounds of exploration.
Again because $\oo{1} < \zz{1}$, none of the first $T-1$ rounds of exploration explore any $(1,1)$-team; thus, each of these rounds, as well as the very first round of the algorithm, incurs a regret of $\oo{1}$.
    
In round $T$, the total regret is the number of $(1,1)$-teams explored in round $T+1$ (because these teams are still played in round $T$). This number is either $(\zo{1} + \oo{1}) \! \mod \zz{1}$ (if \emph{only} $(1,1)$-teams are explored in round $T+1$, then it is the total number of explored teams in round $T+1$), or $\oo{1}$ (if some $(0,1)$-teams are explored in round $T+1$, then \emph{all} $(1,1)$-teams are explored in round $T+1$). Thus, the regret in the $T^{\text{th}}$ round of exploration is the minimum of the two terms. We thus obtain the total regret of the algorithm as: $\floor{\frac{\zo{1} + \oo{1}}{\zz{1}}} \cdot \oo{1} + \min(\oo{1}, (\zo{1} + \oo{1}) \! \mod \zz{1})$.
\end{proof}

\noindent $\zz{t}$ turns out to be further constrained, as follows:

\begin{lemma} \label{lem:zz-lowerbound}
    In~\cref{alg:or4clique}, if the adversary reveals 0-agents only by exploration in rounds $t \geq 2$, then $\zz{1} > \frac{\al}{5} \N$.
\end{lemma}

\begin{proof}
  The number of 0-agents discovered in round 1 is $2\zz{1}$. In rounds 2 and 3 combined, the algorithm discovers an additional $e_2+e_3$ 0-agents. The number of 1-agents discovered in round 2 is $\Diff{1} - e_2 = 2\zz{1} - e_2$, and in round 3, it is $\Diff{2} - e_3 = 2e_2 - e_3$, by Lemma~\ref{lem:delta-properties}. Thus, the number of unknown 0-agents after round 3 is $\al \N - 2\zz{1} - e_2 - e_3$, and the number of unknown 1-agents is $(1-\al) \cdot \N - 2\zz{1} - e_2 + e_3$. But notice also that after round 3, all unknown agents form 4-cliques of successful edges, which can contain at most one 0-agent each. Therefore, there must be at least three times as many remaining 1-agents as 0-agents, so
  $(1-\al) \cdot \N - 2\zz{1} - e_2 + e_3 \geq 3 (\al \N - 2\zz{1} - e_2 - e_3)$.
  Rearranging, we obtain that $4\zz{1} \geq (4\al-1) \cdot \N - 2e_2 - 4e_3$.
  By Lemma~\ref{lem:delta-properties}, we get that $e_3 \leq e_2 \leq \zz{1}$, so the previous inequality in particular implies that
  $4\zz{1} \geq (4\al-1) \cdot \N - 6 \zz{1}$, or $\zz{1} \geq \frac{4\al-1}{10} \cdot \N > \frac{\al}{5} \cdot \N$,
  because $\al > \frac{1}{2}$.
\end{proof}

To conclude, we get the piecewise-linear bound in \Cref{thm:pwlinearub_improved} by maximizing the bound in~Lemma~\ref{lem:regretbound} subject to the constraint in~Lemma~\ref{lem:zz-lowerbound} (and using $\zo{t} = \al \N - 2 \zz{t}$, $\oo{t} = \zz{t} - (\al - \frac{1}{2}) \N$).

Recall that \cref{lem:regretbound} establishes the following regret bound:

\begin{align*}
    R(\zz{},\zo{},\oo{}) := \floor{\tfrac{\zo{} + \oo{}}{\zz{}}} \cdot \oo{} + \min\left(\oo{}, (\zo{} + \oo{}) \! \mod \zz{}\right).
\end{align*}

To obtain a more convenient form, we now rewrite $R$ as a function of only $\zz{}$.

\begin{lemma} \label{lem:R-rewritten}
  The regret bound can be expressed as a function of $\zz{}$ as
  \begin{align*}
    f_{\al}(\zz{})  &:= \floor{\tfrac{\N}{2\zz{}}} \cdot \left( \tfrac{\N}{2} - \al\N \right) + \min \left( \floor{\tfrac{\N}{2\zz{}}} \cdot \zz{}, \al\N - \zz{} \right).
  \end{align*}
\end{lemma}

\begin{proof}
  Using that $\zo{} = \al\N - 2\zz{}$ and $\oo{}= \tfrac{\N}{2} - \al\N + \zz{}$, we can simplify the regret expression as follows:
    
    \begin{align*}
        R(\zz{},\zo{},\oo{})
        &= \floor{\tfrac{\zo{} + \oo{}}{\zz{}}} \cdot \oo{} + \min \left( \oo{}, (\zo{} + \oo{}) \!\mod \zz{} \right) \\
        &= \floor{\tfrac{\N}{2\zz{}} - 1} \cdot \oo{} + \min \left( \oo{}, \tfrac{\N}{2} \!\mod \zz{} \right) 
        \tag{because $\zo{} + \oo{} = \tfrac{\N}{2} - \zz{}$} \\
        &= \floor{\tfrac{\N}{2\zz{}} - 1} \cdot \oo{} + \min \left( \oo{}, \tfrac{\N}{2} - \zz{} \cdot \floor{\tfrac{\N}{2\zz{}}} \right) 
            \tag{because $\tfrac{\N}{2} \!\mod \zz{} = \tfrac{\N}{2} - \zz{} \cdot \big\lfloor \tfrac{\N}{2\zz{}} \big\rfloor$} \\  
        &= \floor{\tfrac{\N}{2\zz{}}} \cdot \oo{} + \min \left( 0, \tfrac{\N}{2} - \oo{} - \zz{} \cdot \floor{\tfrac{\N}{2\zz{}}} \right) \\
        &= \floor{\tfrac{\N}{2\zz{}}} \cdot \left( \tfrac{\N}{2} - \al\N + \zz{} \right) + \min \big( 0, \al\N - \zz{} - \zz{} \cdot \floor{\tfrac{\N}{2\zz{}}} \big) 
            \tag{because $\oo{} = \tfrac{\N}{2} - \al\N + \zz{}$} \\
        &= \floor{\tfrac{\N}{2\zz{}}} \cdot \left( \tfrac{\N}{2} - \al\N \right) + \min \left( \floor{\tfrac{\N}{2\zz{}}} \cdot \zz{}, \al\N - \zz{} \right) \\
        & = f_{\al}(\zz{}).
    \end{align*}
\end{proof}

\begin{figure}[ht]
    \centering
    \includegraphics[width=0.6\textwidth]{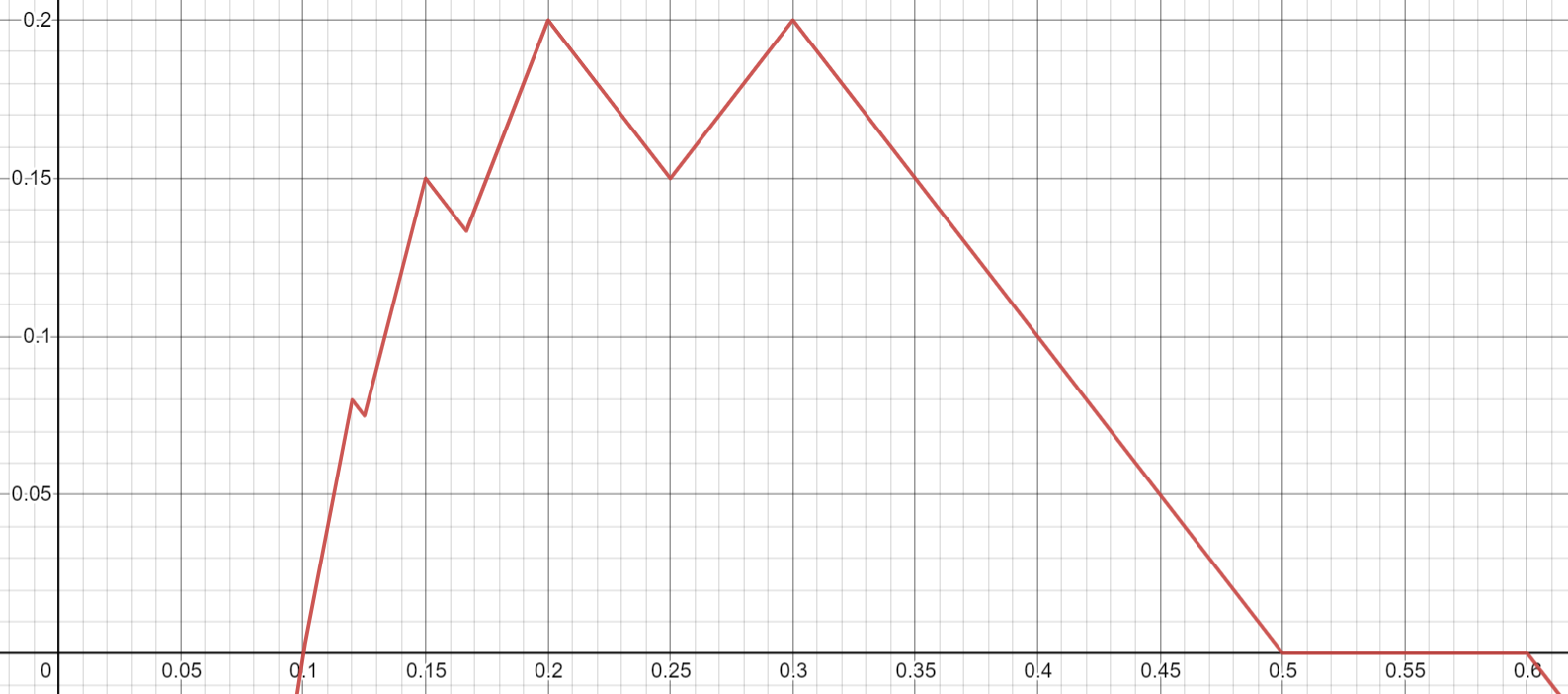}
    \caption{The function $f_{\al}(\zz{})$ for $\al = 0.6$. The local maxima occur at $0.12, 0.15, 0.2$, and $0.3$ (i.e., at $\tfrac{\al}{z}$ for $z\in\{2,3,4,5\}$).}
    \label{fig:saw}
\end{figure}

\begin{lemma} \label{app:lem:xmax}
  For any $\alpha\geq\tfrac{1}{2}$, $R(\zz{},\zo{},\oo{})$ is maximized when $\tfrac{\zz{}}{\N} = \tfrac{\al}{z}$ for some integer $z \geq 2$.
  This implies that
\begin{align}
    R(\zz{},\zo{},\oo{}) \leq \max_{z \in \mathbb{Z}_{>0}}  \left\{ (z-1) \cdot \left( \tfrac{1}{2} - \al + \tfrac{\al}{z} \right) \right\} \cdot \N = \max_{z \in \mathbb{Z}_{>0}}  \left\{ \tfrac{z(z-1) - 2(z-1)^2\al}{2z} \right\} \cdot \N.
\label{eqn:regret-maximum}
\end{align}
\end{lemma}

\begin{proof}
  The function $f_{\al}(\zz{})$ derived in Lemma~\ref{lem:R-rewritten} has a saw-tooth shape (see~\cref{fig:saw}).

  We first prove that at any global maximum of $f_{\al}(\cdot)$, the two terms under the minimum must be equal, i.e., $\big\lfloor \tfrac{\N}{2\zz{}} \big\rfloor \cdot \zz{} =  \al\N - \zz{}$. We do so by distinguishing two cases:
  \begin{enumerate}
  \item If $\frac{\N}{2\zz{}}$ is not an integer, then $\big\lfloor \frac{\N}{x} \big\rfloor$ is constant for $x$ in a small interval around $\zz{}$. In this interval, the first term of the minimum is increasing in $\zz{}$ and the second decreasing in $\zz{}$ (while the term before the minimum is constant in this interval). Therefore, if the two terms under the minimum were not equal, by increasing or decreasing $\zz{}$ slightly, the minimum could be increased.
  \item If $\frac{\N}{2\zz{}}$ is an integer, then we first argue that the minimum must equal the first term. The reason is as follows: when $\zz{}$ is decreased slightly, $\big\lfloor  \frac{\N}{2\zz{}} \big\rfloor$ stays constant, and the $\zz{}$ factor decreases slightly (while the term outside the minimum stays constant). On the other hand, $\al\N - \zz{}$ increases slightly; if the first term were strictly larger than the minimum, then the minimum could be increased by slightly decreasing $\zz{}$.

    On the other hand, when $\frac{\N}{2\zz{}}$ is an integer, the first term under the minimum equals $\frac{\N}{2}$, while the second equals $\al \N - \zz{}$. Because $\al \N$ is the number of 0-agents, $\al \N - \zz{}$ is the number of pairs including at least one 0-agent, which is at most the total number of pairs, $\frac{\N}{2}$. Therefore, the minimum must also equal the second term.
  \end{enumerate}
  
  Thus, we have shown that $\big\lfloor \tfrac{\N}{2\zz{}} \big\rfloor \cdot \zz{} =  \al\N - \zz{}$ at any global maximum of $f_{\alpha}$. 
  Rearranging this equation, we obtain that $\al = \big\lfloor \tfrac{\N}{2\zz{}} + 1\big\rfloor \tfrac{\zz{}}{\N}$ at any local maximum.
  Note that $\zz{} \leq \tfrac{\al\N}{2}$, so $\big\lfloor \tfrac{\N}{2\zz{}} + 1 \big\rfloor$ is an integer, and at least 2.
  Thus, any global maximum occurs when $\tfrac{\zz{}}{\N} = \tfrac{\al}{z}$ for some integer $z \geq 2$. To derive the forms in Inequality~\eqref{eqn:regret-maximum}, we equate the two terms in the minimum expression for $\zz{} = \tfrac{\al\N}{z}$. This gives us that $\floor{\frac{z}{2\al}} \cdot \tfrac{\al\N}{z} = \al\N \cdot \frac{z-1}{z}$, so $\floor{\frac{z}{2\al}} = z-1$. Plugging these expressions into $f_\alpha$, we find that
  \[
    f_\alpha(\tfrac{\al\N}{z}) 
    \,=\, \floor{\tfrac{z}{2\al}} \cdot \left( \tfrac{\N}{2} - \al\N \right) + \floor{\tfrac{z}{2\al}} \cdot \tfrac{\al\N}{z}
    \,=\, \floor{\tfrac{z}{2\al}} \cdot \big( \tfrac{1}{2} - \al + \tfrac{\al}{z} \big) \cdot n
    \,=\, (z-1) \cdot \big( \tfrac{1}{2} - \al + \tfrac{\al}{z} \big) \cdot n.
  \]
  
  \noindent This completes the proof of the lemma.
\end{proof}

\noindent Using Lemma~\ref{lem:zz-lowerbound}, we have the additional constraint that $z \leq 5$.
In this case, \cref{eqn:regret-maximum} simplifies to
\[
    \max  \left\{ 0, \tfrac{1 - \al}{2}, \tfrac{3 - 4\al}{3}, \tfrac{6 - 9\al}{4}, \tfrac{10 - 16\al}{5} \right\} \cdot \N.
\]
By determining which of these linear functions is largest for each $\frac{1}{2} < \al \leq 1$, we obtain the bound from \Cref{thm:pwlinearub_improved}.

%% file: orplot.tex
\begin{tikzpicture}[scale=0.45]
    
        \draw[thick] (0,0) -- (15,0);
        \draw[thick] (0,0) -- (0,8.5);
        
        \foreach\x in {0,0.75,...,14.25} {
            \draw[black!30] (\x,0) -- (\x,8.5);
        }
        \foreach\y in {0,0.5,...,8} {
            \draw[black!30] (0,\y) -- (15,\y);
        }
    
        \node at (-0.25,-0.4) {\scriptsize 0};
        \node at (3.75,-0.4) {\scriptsize 0.25};
        \node at (7.5,-0.4) {\scriptsize 0.5};
        \node at (11.25,-0.4) {\scriptsize 0.75};
        \node at (15,-0.4) {\scriptsize 1};
        
        \node at (-0.5,2) {\scriptsize 0.1};
        \node at (-0.5,4) {\scriptsize 0.2};
        \node at (-0.5,6) {\scriptsize 0.3};
        \node at (-0.5,8) {\scriptsize 0.4};
        
        \node at (7.5, -1.5) { $\alpha$};
        \node at (-2.2,4.25) {\Large $\frac{\textrm{Regret}}{n}$};
        
        \draw[very thick, blue] (0,0) -- (7.5,8);
        \draw[very thick, blue] (7.5,8) -- (150/19,120/19);
        
        \draw[very thick, red] (0,0) -- (7.5,130/17);
        \draw[very thick, red] (7.5,60/8) -- (150/19,120/19);
        
        \draw[very thick, violet] (150/19,120/19) -- (90/11,60/11);
        \draw[very thick, violet] (90/11,60/11) -- (9,4);
        \draw[very thick, violet] (9,4) -- (15,0);
        
        \draw[thick, fill=white] (10,5.25) -- (10,7.75) -- (14.625,7.75) -- (14.625,5.25) -- cycle;
        \draw[very thick, blue] (10.375,7.25) -- (11.125,7.25);
        \node[blue] at (12.25,7.25) {\scriptsize $U^{\textsf{OR}}(\al)$};
        \draw[very thick, red] (10.375,6.5) -- (11.125,6.5);
        \node[red] at (12.25,6.5) {\scriptsize $L^{\textsf{OR}}(\al)$};
        \draw[very thick, violet] (10.375,5.75) -- (11.125,5.75);
        \node[violet] at (12.75,5.75) {\scriptsize $U^{\textsf{OR}} = L^{\textsf{OR}}$};
    \end{tikzpicture}

%% file: s5_and.tex
\section{The Weakest Link Setting (\textsf{AND})} \label{sec:and}

Finally, we consider the Boolean \textsf{AND} synergy function.
If, as before, we interpret 0/1-agents as having low/high skill, then (in the terminology of Johari et al. \cite{Johari18_Team}), this corresponds to a \emph{weakest link} model: the difficulty of the task ensures any team with a low-skill member is unsuccessful. To simplify the analysis, we assume throughout that $\K$ is even.

\begin{theorem} \label{thm:andlb}
    For the weakest link model, any algorithm incurs regret at least $L^{\textsf{AND}}(\N,\K) := \N-\K$.
\end{theorem}


We will prove this theorem by describing a greedy policy of the adversary that establishes the lower regret bound $\N-\K$ for the weakest link model. Note that the undiscovered subgraph for this model consists entirely of unsuccessful edges. Thus, the 1-agents form an independent set in this subgraph. 

A \emph{labeling} is a function $\lambda : [n] \to \{ 0,1 \}$ assigning the skill of each agent. It is \emph{viable} if $\sum_{i=1}^{n} \lambda(i) = k$ (it assigns exactly $k$ 1-agents) and it agrees with all previously revealed edges. 
    
Suppose that $G$ is an undiscovered subgraph on which the algorithm is about to play a matching $M$. Further suppose that there are $j$ undiscovered 1-agents in $G$. Then, a subset $S \subseteq M$ is a \emph{revealable set} if there exists a viable labeling under which the edges in $S$ are successful, and the edges in $M \setminus S$ are unsuccessful.

A \emph{minimal revealable set} is a revealable set for which no proper subset is a revealable set. Note that whenever $|M| \geq \frac{j}{2}$, any $\frac{j}{2}$-subset of $M$ is revealable: all remaining nodes can be 0-agents according to the previously revealed information. Therefore, a \emph{minimal revealable set} exists. Note that if $G \cup M$ has an independent set of size $j$, then $\emptyset$ is the unique minimal revealable set.
With this terminology, we can describe the policy of the adversary in two consecutive steps per round:

\begin{enumerate}
    \item Iterate over the \emph{explored agents} $u$, i.e., the yet-undiscovered agents that are paired with a discovered 1-agent in this round. If revealing $u$ as a 0-agent still admits a viable labeling of the remaining agents, the adversary reveals $u$ as a 0-agent. Otherwise, $u$ is revealed as a 1-agent.  
    \item Let $G$ denote the resulting undiscovered graph after labeling all nodes whose type could be deduced from step 1. The adversary chooses a minimal revealable set $S$ for $G$, labels these edges as successful, and the remaining edges as unsuccessful.
\end{enumerate}

To argue the regret bound, we make use of the following three lemmas.

\begin{lemma} \label{lem:exp2connect}
    Let $u$ be a 1-agent explored by being paired with a known 1-agent. Let $w$ be a 0-agent that was discovered through its connection to $u$. Then, in all viable labelings, $w$ is connected to at least two 1-agents.
\end{lemma}

\begin{proof}
    For the sake of contradiction, suppose that there is a viable labeling $\lambda$ in which $w$'s only 1-agent connection is to $u$. Consider the alternative labeling $\lambda'$ which switches the labels of $u$ and $w$ in $\lambda$, but keeps all other labels the same. This labeling is also viable at this point since neither $w$ (by assumption) nor $u$ (it is undiscovered) has been paired with any other 1-agent. However, the existence of $\lambda'$ contradicts the adversary's necessity to reveal $u$ as a 1-agent. 
\end{proof}

\begin{lemma} \label{lem:disc2connect}
  Let $G$ be an undiscovered subgraph from step 2 containing $j$ 1-agents. Let $S$ be a minimal revealable set, and let $w$ be a 0-agent that was discovered through its connection to an endpoint $u$ of an edge in $S$. Then, in all viable labelings (assigning type 1 to all%
  \footnote{Note that no edge in $S$ can have been played in a previous round --- otherwise, revealing it as a successful edge would not leave any viable labelings, as it would contradict the earlier revelation as unsuccessful. This would contradict the definition of $S$ being revealable.}
  endpoints of edges in $S$), $w$ is connected to at least two 1-agents.
\end{lemma}

\begin{proof}
    For the sake of contradiction, suppose that there is a labeling $\lambda$ for which the only 1-agent that $w$ is connected to is $u$. Let $v$ denote the other endpoint of $u$'s revealed 1-edge (so $(u,v) \in S$). Finally, let $I$ denote a set of $j-2|S|$ 1-agents (given by $\lambda$) in the resulting undiscovered subgraph (guaranteed to exist because $S$ is revealable). By assumption, $w$ is connected to neither $v$ nor any agent in $I$. Additionally, no agent in $I$ is connected to $v$ (because the agents in $I$ are undiscovered after $(u,v)$ is revealed as a 1-edge). Therefore, $I \cup \{ w,v \}$ is an independent set of size $j-2|S|+2$ in $G$. However, this means that $S \setminus \{ (u,v) \}$ is a revealable set (a viable labeling $\lambda'$ is obtained by starting with $\lambda$, and switching the label of $u$ to 0 while switching the label of $w$ to 1), contradicting the minimality of $S$.  
\end{proof}

\begin{lemma} \label{lem:everyoneconnected}
    Suppose that the adversary (using this policy) has just revealed the first 1-edge(s). Then, in any viable labeling $\lambda$, each undiscovered 0-agent is connected to at least one undiscovered 1-agent. 
\end{lemma}

\begin{proof}
  Let $S$ be the minimal revealable set that was just revealed by the adversary. Fix any viable labeling $\lambda$, and let $I$ denote the set of undiscovered 1-agents in $\lambda$. For the sake of contradiction, suppose that there is an undiscovered 0-agent $w$ that is not paired with an agent in $I$. Then, $I \cup \{w\}$ is an independent set of size $j-2|S| + 1$. If $(u,v) \in S$ is any successful edge that was just revealed, then $u$ has no edges to $I$ (by viability of $\lambda$) nor to $w$ (or $w$ would have been discovered as a 0-agent). Therefore, $I \cup \{w,u\}$ is an independent set of size $j-2|S|+2$, so $S \setminus \{ (u,v) \}$ is a revealable set; the viable labeling $\lambda'$ witnessing this is obtained from $\lambda$ by switching the label of $w$ to 1 and the label of $v$ to 0.
  This contradicts the minimality of $S$.
\end{proof}

Using these lemmas, we can show that any policy incurs a regret of at least $n-k$ under this adversarial policy. 

\begin{extraproof}{\Cref{thm:andlb}}
  It suffices to argue that, at the time of its discovery, each 0-agent must have been paired with at least two 1-agents in any viable labeling. (Lemma~\ref{lem:everyoneconnected} guarantees that all 0-agents are connected to a 1-agent after the first discovery, so once all 1-agents are discovered, all 0-agents will also have been discovered.) Lemmas~\ref{lem:exp2connect} and~\ref{lem:disc2connect} ensure this for all 0-agents that are discovered by having a connection to a 1-agent that is discovered. Lemma~\ref{lem:everyoneconnected} ensures this for all 0-agents $u$ that are explored by a known 1-agent $w$. This is because $u$ will be connected to $w$ and the guaranteed undiscovered 1-agent $v$, which it was connected to earlier (implying that $v \neq w$).

  Therefore, the algorithm plays at least $2(n-k)$ $(0,1)$-edges, leading to regret at least $n-k$.
\end{extraproof}

\subsection{The \ringalg Algorithm} \label{subsec:ringalg}

The fact that the regret of an algorithm is half the number of $(0,1)$-teams selected suggests that we want the algorithm's chosen matchings to quickly locate (and pair) all of the 1-agents, while minimizing the number of times each 0-agent is paired with a 1-agent. Playing matchings according to a \emph{1-factorization} (that is, a partition of the complete graph $K_\N$ into perfect matchings) ensures that no team is ever repeated. This intuition is used in the \textsc{Exponential Cliques} algorithms of Johari et al. \cite{Johari18_Team}, who show that when each agent has independent Bernoulli($\K/\N$) type, this algorithm has expected regret $\frac{3}{4}(\N - \K) + o(\N)$, which is asymptotically optimal.
Against an adaptive adversary, however, an arbitrary 1-factorization is not enough to get good regret; for example, a 1-factorization that first builds the Tur\'{a}n graph $T(\N,\frac{\N}{\K})$ \cite{Turan41,Aigner95_Turan} has regret $\frac{1}{2} \K(\N-\K)$. 
Similarly, the performance of \textsc{Exponential Cliques} in the worst case is also much worse.
  
\begin{lemma}
\label{lem:expcliques}
\textsc{Exponential Cliques} incurs regret $2(\N-\K-1)$ against an adaptive adversary. 
\end{lemma}

\begin{proof}
    Consider an instance on $\N = 2^j+2$ agents with $\K=2$ having high skill. An adaptive adversary can ensure that the two 1-agents comprise the last unexplored team. Over its first $2^j-1$ rounds, Exponential Cliques builds a $2^j$-clique in the explored subgraph while repeating the remaining team $2^j-1$ times. Subsequently, it must spend $2^j$ additional rounds exploring all teams comprising a member of this repeated edge and a member of the clique, resulting in regret $2(2^j-1) = 2(\N-\K-1)$.
\end{proof}

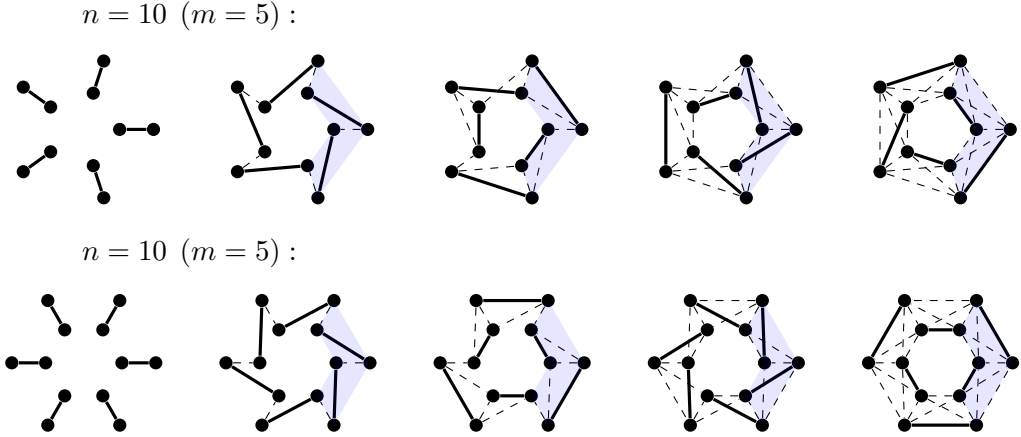
\begin{figure*}[htb]
    \centering
    \import{./}{ringfig.tex}
    \caption{\small\em The first five rounds (i.e., Phases 0 and 1) of \ringfac on 10 (top) and 12 (bottom) agents. The last four matchings illustrate the general matching sequence for cycles in intermediate phases; the blue highlighted section of each matching is repeated based upon the size of the cycle. 
    }
    \label{fig:1012fact}
\end{figure*}

Our main algorithm for this setting, \ringalg, leverages a particular 1-factorization, which we call the \ringfac. We organize the agents into two nested rings, and choose matchings so that closer agent pairs under this ring geometry are matched earlier. In the first round, agents in corresponding positions in the rings are matched. Over the next four rounds, matchings are chosen to pair each agent with the four agents in adjacent positions to it in the rings, and this process repeats at greater distances. The structure and order of the four matchings chosen in each ``phase'' are critical. A formal description of this 1-factorization is given in~\cref{app:ringfact}, and we visualize the first 5 matchings in the construction for $n=10$ and $n=12$ in~\cref{fig:1012fact}.

\begin{algorithm}[htb]
    \caption{\ringalg (Sketch)}
    \label{alg:ring}
    \begin{algorithmic} 
        \WHILE { $\# \{ \text{unknown agents} \} > 0$}
            \STATE Select a matching via \ringfac for $\N$.
            \IF{ a $(1,1)$-team is revealed }
            \STATE Perform a case-specific ``repair'' step (possibly over multiple rounds) that partitions agents into known $(0,0)$-teams, known $(1,1)$-teams, and an intermediary stage of the \ringfac construction of size $\N' < \N$. (See~\cref{app:ringalg}.)
            \STATE Play known $(0,0)$- and $(1,1)$-teams, and continue playing matchings according to \ringfac on the remaining $\N'$ agents.
            \ENDIF
        \ENDWHILE
    \end{algorithmic}
\end{algorithm}

\begin{theorem} \label{thm:andub}
    \ringalg (\cref{alg:ring}) locates an optimal matching after incurring regret at most $U^{\textsf{AND}}(\N,\K) := \N - \K + \floor{\tfrac{\min(\K,\N-\K)}{4}}.$
\end{theorem}

\begin{proofsketch}
    As mentioned before, the double-ring structure of our factorization defines a notion of distance between agents (namely, the difference between their column indices modulo $m$). By selecting matchings according to this factorization, each agent is paired up with other agents in non-decreasing order of distance. Consider this pairing from the perspective of a 0-agent $x$. We will show that roughly speaking (with some exceptions which require technical work and slightly weaken the bound), $x$ will be paired with at most two 1-agents before being identified as a 0-agent. If each 0-agent is paired with at most two 1-agents before discovery, then we get an overall regret bound of $\N-\K$. Consider three 1-agents $\{ y_1, y_2, y_3 \}$, all located in different columns from $x$. Then, two of these 1-agents --- say, $y_1$ and $y_2$ --- lie on the same side of $x$. Thus, $y_1$ and $y_2$ are strictly closer to each other than the further of the two (say, $y_1$) is to $x$. In particular, $y_1$ and $y_2$ were paired before $y_1$ is paired with $x$, and so must have been revealed as 1-agents. Thus, $y_1$ will never be paired with $x$. Since this holds for every triple of 1-agents, $x$ cannot be paired with three 1-agents.
    
    While the above argument encapsulates the main intuition, the technical challenge is removing the assumption that $y_1, y_2, y_3$ were all in different columns from $x$. ``Repairing'' the cycle to account for the case of a 1-agent in the same column as $x$ largely accounts for the additional term in the regret bound. The full analysis of these ``repair'' steps is intricate; see \cref{app:ringalg} for details. 
\end{proofsketch}

\noindent The bounds of Theorems~\ref{thm:andlb} and~\ref{thm:andub} are off by an additive term $\floor{\min(\K,\N-\K) / 4}$. 
The lower bound is simpler, and it is tempting to think that it may be tight; unfortunately, this is not true in general; in~\cref{app:104}, we show that any algorithm on the instance with $\N=10$, $\K=4$ must incur regret at least 7 (which coincidentally matches the upper bound in Theorem~\ref{thm:andub}, though it is unclear if this extends to larger settings).
Closing this gap is an interesting and challenging direction for future work.

%% file: ringfig.tex
$\N=10 \:\:(m=5): \hspace{250pt}$ \\[8pt]

\begin{minipage}{0.14\textwidth}
\begin{tikzpicture}[scale=0.29]
    \def\n{5}
   \node[circle,minimum size=1.9cm] (a) {};
    \node[circle,minimum size=1 cm] (b) {};
    \foreach\x in{1,...,\n}{
      \node[inner sep=0pt,minimum size=0.175cm,fill,circle] (a\x) at (a.{360/\n*\x}){};
      \node[inner sep=0pt,minimum size=0.175cm,fill,circle] (b\x) at (b.{360/\n*\x}){};
    }
    \draw[very thick] (a1) -- (b1) (a2) -- (b2) (a3) -- (b3) (a4) -- (b4) (a5) -- (b5);
\end{tikzpicture}
\end{minipage}
\hspace{12pt}
\begin{minipage}{0.14\textwidth}
\begin{tikzpicture}[scale=0.29]
    \def\n{5}
   \node[circle,minimum size=1.9cm] (a) {};
    \node[circle,minimum size=1 cm] (b) {};
    \foreach\x in{1,...,\n}{
      \node[inner sep=0pt,minimum size=0.175cm,fill,circle] (a\x) at (a.{360/\n*\x}){};
      \node[inner sep=0pt,minimum size=0.175cm,fill,circle] (b\x) at (b.{360/\n*\x}){};
    }
    \draw[dashed] (a1) -- (b1) (a2) -- (b2) (a3) -- (b3) (a4) -- (b4) (a5) -- (b5);
    \draw[very thick] (a1) -- (b2) (a2) -- (b3) (a3) -- (b4) (a4) -- (b5) (a5) -- (b1);
    
    \fill[blue, opacity=0.1] (b1.center) -- (b5.center) -- (b4.center) -- (a4.center) -- (a5.center) -- (a1.center) -- cycle;
\end{tikzpicture}
\end{minipage}
\hspace{12pt}
\begin{minipage}{0.14\textwidth}
\begin{tikzpicture}[scale=0.29]
    \def\n{5}
   \node[circle,minimum size=1.9cm] (a) {};
    \node[circle,minimum size=1 cm] (b) {};
    \foreach\x in{1,...,\n}{
      \node[inner sep=0pt,minimum size=0.175cm,fill,circle] (a\x) at (a.{360/\n*\x}){};
      \node[inner sep=0pt,minimum size=0.175cm,fill,circle] (b\x) at (b.{360/\n*\x}){};
    }
    \draw[dashed] (a1) -- (b1) (a2) -- (b2) (a3) -- (b3) (a4) -- (b4) (a5) -- (b5);
    \draw[dashed] (a1) -- (b2) (a2) -- (b3) (a3) -- (b4) (a4) -- (b5) (a5) -- (b1);
    \draw[very thick] (a2) -- (b1) (a3) -- (a4) (a5) -- (a1) (b2) -- (b3) (b4) -- (b5);
    
    \fill[blue, opacity=0.1] (b1.center) -- (b5.center) -- (b4.center) -- (a4.center) -- (a5.center) -- (a1.center) -- cycle;
\end{tikzpicture}
\end{minipage}
\hspace{12pt}
\begin{minipage}{0.14\textwidth}
\begin{tikzpicture}[scale=0.29]
    \def\n{5}
   \node[circle,minimum size=1.9cm] (a) {};
    \node[circle,minimum size=1 cm] (b) {};
    \foreach\x in{1,...,\n}{
      \node[inner sep=0pt,minimum size=0.175cm,fill,circle] (a\x) at (a.{360/\n*\x}){};
      \node[inner sep=0pt,minimum size=0.175cm,fill,circle] (b\x) at (b.{360/\n*\x}){};
    }
    \draw[dashed] (a1) -- (b1) (a2) -- (b2) (a3) -- (b3) (a4) -- (b4) (a5) -- (b5);
    \draw[dashed] (a1) -- (b2) (a2) -- (b3) (a3) -- (b4) (a4) -- (b5) (a5) -- (b1);
    \draw[dashed] (a2) -- (b1) (a3) -- (a4) (a5) -- (a1) (b2) -- (b3) (b4) -- (b5);
    \draw[very thick] (b1) -- (b2) (a2) -- (a3) (b3) -- (a4) (b4) -- (a5) (b5) -- (a1);
    
    \fill[blue, opacity=0.1] (b1.center) -- (b5.center) -- (b4.center) -- (a4.center) -- (a5.center) -- (a1.center) -- cycle;
\end{tikzpicture}
\end{minipage}
\hspace{12pt}
\begin{minipage}{0.14\textwidth}
\begin{tikzpicture}[scale=0.29]
    \def\n{5}
   \node[circle,minimum size=1.9cm] (a) {};
    \node[circle,minimum size=1 cm] (b) {};
    \foreach\x in{1,...,\n}{
      \node[inner sep=0pt,minimum size=0.175cm,fill,circle] (a\x) at (a.{360/\n*\x}){};
      \node[inner sep=0pt,minimum size=0.175cm,fill,circle] (b\x) at (b.{360/\n*\x}){};
    }
    \draw[dashed] (a1) -- (b1) (a2) -- (b2) (a3) -- (b3) (a4) -- (b4) (a5) -- (b5);
    \draw[dashed] (a1) -- (b2) (a2) -- (b3) (a3) -- (b4) (a4) -- (b5) (a5) -- (b1);
    \draw[dashed] (a2) -- (b1) (a3) -- (a4) (a5) -- (a1) (b2) -- (b3) (b4) -- (b5);
    \draw[dashed] (b1) -- (b2) (a2) -- (a3) (b3) -- (a4) (b4) -- (a5) (b5) -- (a1);
    \draw[very thick] (b2) -- (a3) (a4) -- (a5) (a1) -- (a2) (b3) -- (b4) (b5) -- (b1);
    
    \fill[blue, opacity=0.1] (b1.center) -- (b5.center) -- (b4.center) -- (a4.center) -- (a5.center) -- (a1.center) -- cycle;
\end{tikzpicture}
\end{minipage}

\vspace{12pt}
$\N=10 \:\:(m=5): \hspace{250pt}$ \\[8pt]

\begin{minipage}{0.14\textwidth}
\begin{tikzpicture}
    \def\n{6}
    \node[circle,minimum size=1.9cm] (a) {};
    \node[circle,minimum size=1 cm] (b) {};
    \foreach\x in{1,...,\n}{
      \node[inner sep=0pt,minimum size=0.175cm,fill,circle] (a\x) at (a.{360/\n*\x}){};
      \node[inner sep=0pt,minimum size=0.175cm,fill,circle] (b\x) at (b.{360/\n*\x}){};
    }
    
    \draw[very thick] (a1) -- (b1) (a2) -- (b2) (a3) -- (b3) (a4) -- (b4) (a5) -- (b5) (a6) -- (b6);
\end{tikzpicture}
\end{minipage}
\hspace{12pt}
\begin{minipage}{0.14\textwidth}
\begin{tikzpicture}
    \def\n{6}
    \node[circle,minimum size=1.9cm] (a) {};
    \node[circle,minimum size=1 cm] (b) {};
    \foreach\x in{1,...,\n}{
      \node[inner sep=0pt,minimum size=0.175cm,fill,circle] (a\x) at (a.{360/\n*\x}){};
      \node[inner sep=0pt,minimum size=0.175cm,fill,circle] (b\x) at (b.{360/\n*\x}){};
    }
    
    \draw[dashed] (a1) -- (b1) (a2) -- (b2) (a3) -- (b3) (a4) -- (b4) (a5) -- (b5) (a6) -- (b6);
    \draw[very thick] (a1) -- (b2) (a2) -- (b3) (a3) -- (b4) (a4) -- (b5) (a5) -- (b6) (a6) -- (b1);
    
    \fill[blue, opacity=0.1] (b1.center) -- (b6.center) -- (b5.center) -- (a5.center) -- (a6.center) -- (a1.center) -- cycle;
\end{tikzpicture}
\end{minipage}
\hspace{12pt}
\begin{minipage}{0.14\textwidth}
\begin{tikzpicture}
    \def\n{6}
    \node[circle,minimum size=1.9cm] (a) {};
    \node[circle,minimum size=1 cm] (b) {};
    \foreach\x in{1,...,\n}{
      \node[inner sep=0pt,minimum size=0.175cm,fill,circle] (a\x) at (a.{360/\n*\x}){};
      \node[inner sep=0pt,minimum size=0.175cm,fill,circle] (b\x) at (b.{360/\n*\x}){};
    }
    
    \draw[dashed] (a1) -- (b1) (a2) -- (b2) (a3) -- (b3) (a4) -- (b4) (a5) -- (b5) (a6) -- (b6);
    \draw[dashed] (a1) -- (b2) (a2) -- (b3) (a3) -- (b4) (a4) -- (b5) (a5) -- (b6) (a6) -- (b1);
    \draw[very thick] (a1) -- (a2) (b2) -- (b3) (a3) -- (a4) (b4) -- (b5) (a5) -- (a6) (b6) -- (b1);
    
    \fill[blue, opacity=0.1] (b1.center) -- (b6.center) -- (b5.center) -- (a5.center) -- (a6.center) -- (a1.center) -- cycle;
\end{tikzpicture}
\end{minipage}
\hspace{12pt}
\begin{minipage}{0.14\textwidth}
\begin{tikzpicture}
    \def\n{6}
    \node[circle,minimum size=1.9cm] (a) {};
    \node[circle,minimum size=1 cm] (b) {};
    \foreach\x in{1,...,\n}{
      \node[inner sep=0pt,minimum size=0.175cm,fill,circle] (a\x) at (a.{360/\n*\x}){};
      \node[inner sep=0pt,minimum size=0.175cm,fill,circle] (b\x) at (b.{360/\n*\x}){};
    }
    
    \draw[dashed] (a1) -- (b1) (a2) -- (b2) (a3) -- (b3) (a4) -- (b4) (a5) -- (b5) (a6) -- (b6);
    \draw[dashed] (a1) -- (b2) (a2) -- (b3) (a3) -- (b4) (a4) -- (b5) (a5) -- (b6) (a6) -- (b1);
    \draw[dashed] (a1) -- (a2) (b2) -- (b3) (a3) -- (a4) (b4) -- (b5) (a5) -- (a6) (b6) -- (b1);
    \draw[very thick] (b1) -- (a2) (b2) -- (a3) (b3) -- (a4) (b4) -- (a5) (b5) -- (a6) (b6) -- (a1);
    
    \fill[blue, opacity=0.1] (b1.center) -- (b6.center) -- (b5.center) -- (a5.center) -- (a6.center) -- (a1.center) -- cycle;
\end{tikzpicture}
\end{minipage}
\hspace{12pt}
\begin{minipage}{0.14\textwidth}
\begin{tikzpicture}
    \def\n{6}
    \node[circle,minimum size=1.9cm] (a) {};
    \node[circle,minimum size=1 cm] (b) {};
    \foreach\x in{1,...,\n}{
      \node[inner sep=0pt,minimum size=0.175cm,fill,circle] (a\x) at (a.{360/\n*\x}){};
      \node[inner sep=0pt,minimum size=0.175cm,fill,circle] (b\x) at (b.{360/\n*\x}){};
    }
    
    \draw[dashed] (a1) -- (b1) (a2) -- (b2) (a3) -- (b3) (a4) -- (b4) (a5) -- (b5) (a6) -- (b6);
    \draw[dashed] (a1) -- (b2) (a2) -- (b3) (a3) -- (b4) (a4) -- (b5) (a5) -- (b6) (a6) -- (b1);
    \draw[dashed] (a1) -- (a2) (b2) -- (b3) (a3) -- (a4) (b4) -- (b5) (a5) -- (a6) (b6) -- (b1);
    \draw[dashed] (b1) -- (a2) (b2) -- (a3) (b3) -- (a4) (b4) -- (a5) (b5) -- (a6) (b6) -- (a1);
    \draw[very thick] (b1) -- (b2) (a2) -- (a3) (b3) -- (b4) (a4) -- (a5) (b5) -- (b6) (a6) -- (a1);
    
    \fill[blue, opacity=0.1] (b1.center) -- (b6.center) -- (b5.center) -- (a5.center) -- (a6.center) -- (a1.center) -- cycle;
\end{tikzpicture}
\end{minipage}

%% file: sa4_ring_details.tex
\section{\ringfac: Detailed Construction} \label{app:ringfact}

Let $m := \frac{\N}{2}$, and arbitrarily label the $\N$ agents as $u_0, u_1, \hdots, u_{m-1}$ and $v_0, v_1, \hdots$, $v_{m-1}$. In \cref{fig:1012fact}, the $u_i$ are the nodes on the inner ring; the pair of vertices with matching indices form a ``column'' of the ring. We separate the factorization into $\ceil{\frac{m+1}{2}}$ phases (numbered starting from 0). In each phase $i$, we select matchings to exactly add the edges connecting agents whose subscripts are $i$ apart (modulo $m$). Note that agent subscripts differ by at most $\lfloor \frac{m}{2} \rfloor$, so this accounts for all edges.

{
\setlength{\parskip}{6pt}
\setlength{\parindent}{0pt}

\vspace{4pt}
\noindent\textbf{Phase 0:} This phase lasts for 1 round

Select the edges $\{(u_0,v_0),(u_1,v_1),\hdots,(u_{m-1},v_{m-1})\}$. 

\vspace{10pt}
\noindent\textbf{Phase $\bm{i} \hspace{5pt} \bm{(1 \!\leq\! i \!<\! \frac{m}{2})}$:} Each of these phases lasts 4 rounds.

\parshape 1 \parindent \dimexpr\linewidth-\parindent
\noindent  Consider the graph on $[m]$ with an edge between two indices iff they are exactly $i$ apart modulo $m$. This graph is a union of $\gcd(m,i)$ disjoint cycles, each containing $j := \frac{m}{\gcd(m,i)} \geq 3$ indices.

\parshape 1 \parindent \dimexpr\linewidth-\parindent
\noindent It suffices to define a procedure for selecting perfect matchings within each of these cycles.
Focus on one cycle of length $j$, and relabel the corresponding $2j$ agents as $u'_0, u'_1, \hdots, u'_{j-1}$ and $v'_0, v'_1, \hdots, v'_{j-1}$. There are two cases based on whether $j$ is even or odd.

If $j$ is even, the factorization uses the following four matchings:

\hspace{\parindent} \textbf{Round 1:} \: $\Big\{ (u'_0, v'_1),(u'_1, v'_2), \hdots, (u'_{j-1}, v'_0) \Big\}$

\hspace{\parindent} \textbf{Round 2:} \: $\Big\{ (u'_0, u'_1),(u'_2, u'_3), \hdots, (u'_{j-2}, u'_{j-1}) \Big\} \, \cup \, \Big\{ (v'_1, v'_2),(v'_3, v'_4), \hdots, (v'_{j-1}, v'_0) \Big\}$

\hspace{\parindent} \textbf{Round 3:} \: $\Big\{ (v'_0, u'_1),(v'_1, u'_2), \hdots, (v'_{j-1}, u'_0) \Big\}$

\hspace{\parindent} \textbf{Round 4:} \: $\Big\{ (u'_1, u'_2),(u'_3, u'_4), \hdots, (u'_{j-1}, u'_0) \Big\} \, \cup \, \Big\{ (v'_0, v'_1),(v'_2, v'_3), \hdots, (v'_{j-2}, v'_{j-1}) \Big\}$

If $j$ is odd, the factorization uses the following four matchings:

\hspace{\parindent} \textbf{Round 1:} \: $\Big\{ (u'_0, v'_1), (u'_1, v'_2), \hdots, (u'_{j-1}, v'_0) \Big\}$

\hspace{\parindent} \textbf{Round 2:} \: $\Big\{ (v'_0, u'_1) \Big\} \, \cup \, \Big\{ (u'_2, u'_3),(u'_4, u'_5), \hdots, (u'_{j-1}, u'_0) \Big\} \, \cup \, \Big\{ (v'_1, v'_2),(v'_3, v'_4), \hdots, (v'_{j-2}, v'_{j-1}) \Big\}$

\hspace{\parindent} \textbf{Round 3:} \: $\Big\{ (v'_0, v'_1), (u'_1, u'_2) \Big\} \, \cup \, \Big\{ (v'_2, u'_3), (v'_3, u'_4), \hdots, (v'_{j-1}, u'_0) \Big\}$

\hspace{\parindent} \textbf{Round 4:} \: $\Big\{ (v'_1, u'_2), (u'_0, u'_1) \Big\} \, \cup \, \Big\{ (u'_3, u'_4), \hdots, (u'_{j-2}, u'_{j-1}) \Big\} \, \cup \, \Big\{ (v'_2, v'_3),(v'_4, v'_5), \hdots, (v'_{j-1}, v'_0) \Big\}$

\parshape 1 \parindent \dimexpr\linewidth-\parindent
\noindent Since each node is part of exactly one cycle, taking the union over all cycles gives a perfect matching in each of the four rounds. 

\noindent\textbf{Phase $\bm{\frac{m}{2}}$ (if $\bm{m}$ is even):} This phase lasts 2 rounds.

\hspace{\parindent} \textbf{Round 1:} \: $\Big\{ (u_0, u_{\frac{m}{2}}), (u_1, u_{\frac{m}{2}+1}), \hdots, (u_{\frac{m}{2}-1}, u_{m-1}) \Big\} \, \cup \, \Big\{ (v_0, v_{\frac{m}{2}}), (v_1, v_{\frac{m}{2}+1}), \hdots, (v_{\frac{m}{2}-1}, v_{m-1}) \Big\}$

\hspace{\parindent} \textbf{Round 2:} \: $\Big\{ (u_0, v_{\frac{m}{2}}), (u_1, v_{\frac{m}{2}+1}), \hdots, (u_{\frac{m}{2}-1}, v_{m-1}) \Big\} \, \cup \, \Big\{ (v_0, u_{\frac{m}{2}}), (v_1, u_{\frac{m}{2}+1}), \hdots, (v_{\frac{m}{2}-1}, u_{m-1}) \Big\}$
}

%% file: sa5_ring_proof.tex
\section{Details of the \ringalg~Algorithm} \label{app:ringalg}

Here, we present a proof of~\Cref{thm:andub}, along with the missing details of the algorithm \ringalg.

We first note that it suffices to prove a regret bound of $\N - \K + \floor{\frac{\K}{4}}$. 
For suppose that $\K > \frac{\N}{2}$. In round 1, there can be at most $\N-\K$ unsuccessful teams, meaning that some $(1,1)$-teams must be revealed. An optimal matching then consists of these initially revealed $(1,1)$-teams, plus an optimal matching on the unresolved agents. Since the unresolved agents were each part of an unsuccessful team in round 1, at least half of them are 0-agents. Notably, the number of 1-agents that remain is at most $\N-\K$. Of course, the initial $(1,1)$-teams did not contribute any regret, so the instance has been reduced to a smaller $(\N',\K')$-instance, with $\N' - \K' = \N - \K$ and $\K' \leq \N - \K$. If we can obtain a regret bound of $\N' - \K' + \floor{\frac{\K'}{4}}$, then for the original instance, it results in a regret bound of
\[
    \N' - \K' + \floor{\frac{\K'}{4}} \leq \N - \K + \floor{\frac{\N-\K}{4}}.
\]

Recall that all the regret incurred by an algorithm can be charged to the $(0,1)$-pairs that it plays. More precisely, any two $(0,1)$-pairs played in a round could be re-paired into a $(0,0)$-pair and a $(1,1)$-pair, which would contain one successful pair instead of zero.
Therefore, we can charge half a unit of regret to each $(0,1)$-pair played by an algorithm.
Hence, the total regret is bounded by

\begin{align}
    \textrm{Regret} 
    & \leq \frac{1}{2} \cdot \textrm{\# \{explored $(0,1)$-pairs\}}
    \; = \; \frac{1}{2} \sum_{\textrm{1-agent } v} \# \left\{ \textrm{$(0,1)$-teams including $v$} \right\}.
\label{eqn:regret-as-0-1}
\end{align}

We will use this viewpoint, and analyze the regret for each 1-agent separately, by analyzing how many 0-agents it was paired with. As described in the proof sketch, ideally, we would like to prove that each 1-agent is paired only with two 0-agents. This is not quite true, and we need a more intricate argument to obtain the (slightly weaker) bound of the theorem.
  
The sketch of the \ringalg Algorithm exhibits a natural recursive structure, making it amenable to an inductive argument. In particular, we argue by induction on $\N$ that \ringalg on $\N$ agents ($\N$ even) with $\K \leq \N$ 1-agents ($\K$ even) locates an optimal matching after incurring regret at most $\N - \K + \floor{\frac{\K}{4}}$. Along the way, we flesh out the details of the repairing step.

The base case $\N \in \{0,2\}$ is trivial, as there is a unique matching that incurs regret zero.
The inductive hypothesis states that for some $\N \geq 4$, for every even $\N' < \N$ and every even $\K'$ with $0 \leq \K' \leq \N'$, the \ringalg Algorithm on $\N'$ agents, $\K'$ of whom are 1-agents, determines an optimal matching after incurring regret at most $\N' - \K' + \floor{\frac{\K'}{4}}$. 

For the inductive step, consider an instance with $\N$ agents, $\K$ of whom are 1-agents. If $\K \leq 1$, then any matching procedure incurs no regret. In particular, the \ringalg algorithm satisfies the upper regret bound. Therefore, it suffices to reason about instances with $\K \geq 2$. Since the \ringalg algorithm initially selects matchings according to the ring factorization (a 1-factorization), it must in some round play an edge between two 1-agents for the first time. We perform a case analysis based on the time of this first 1-edge discovery below.

The inductive step relies on a slightly nuanced way to calculate the total regret of the procedure. Suppose that after round $r$, the algorithm is able to remove $z$ discovered 0-agents (by pairing them up permanently) and $w$ 1-agents (also pairing them up permanently), with the following additional properties:

\begin{itemize}
    \item By adding edges which can be deduced to be unsuccessful from the observed outcomes, the induced subgraph on the remaining agents can be made isomorphic to an intermediary stage of a smaller ring factorization.
    \item There is at most one discovered 1-agent that is not removed.
\end{itemize}


The 1-agents that the algorithm removes will never again be paired with a 0-agent, so they have already accumulated their share of the regret. The remaining $\K-w$ 1-agents have cumulatively been incident on $r(\K-w) - x$ (0,1)-edges, where $x \in \{0,1\}$ is the number of removed 1-agents that the discovered, un-removed 1-agent had been paired with (if there was such an agent).
\ringalg will ensure that in the sub-factorization, the $\K-w$ 1-agents will each be incident on exactly $r$ $(0,1)$-edges.
Most of the technical work of the proof goes into proving the following lemma, which we will do below.

\begin{lemma} \label{lem:removal-bound}
  Each removal of the form described above satisfies

\begin{align} \label{eq3}
  \frac{1}{2} \cdot \sum_{\substack{\textrm{ removed } \\ \textrm{1-agent } v}} \# \left\{ \textrm{$(0,1)$-teams including $v$} \right\}
  & \leq z + \left\lfloor \frac{w}{4} \right\rfloor + \frac{x}{2}. 
\end{align}
\end{lemma}

For the remaining regret, we apply the induction hypothesis to the strictly smaller ring factorization on the remaining agents.
The induction hypothesis guarantees a regret of
\begin{align*}
    \textrm{Regret of Sub-Factorization}
    & \leq (\N - (z+w)) - (\K-w) + \floor{\frac{\K-w}{4}} \\
    & \leq \N-\K + \floor{\frac{\K}{4}} - z - \floor{\frac{w}{4}}.
\end{align*}

Using the bounds from Lemma~\ref{lem:removal-bound} and the induction hypothesis, we can now decompose and bound the regret from Equation~\eqref{eqn:regret-as-0-1} as follows:
\begin{align*}
    \textrm{Regret} 
  & = \tfrac{1}{2} \hspace{-10pt} \sum_{\substack{\textrm{ removed } \\ \textrm{1-agent } v}} \hspace{-8pt} \# \left\{ \textrm{$(0,1)$-teams including $v$} \right\}
    + \textrm{Regret of Sub-Factorization} - \frac{x}{2} \\
  & \leq \left( z + \floor{\frac{w}{4}} + \frac{x}{2} \right) 
    + \left( \N-\K + \floor{\frac{\K}{4}} - z - \floor{\frac{w}{4}} \right)
    - \frac{x}{2} \\
  & = \N-\K + \floor{\frac{\K}{4}}.
\end{align*}
This completes the proof of the induction step, and thus the theorem.

\medskip

In the remainder of this section, we prove Lemma~\ref{lem:removal-bound}.
\begin{extraproof}{Lemma~\ref{lem:removal-bound}}
  We perform a detailed case analysis based on the time of the first discovery of a successful edge. We always denote the successful edge by $(a,b)$.

\begin{enumerate}
\item Discovery during phase 0: 

Removing the endpoints of the discovered edge gives an intermediary stage of the sub-factorization on $\N-2$ agents with $(\K-2)$ 0-agents. In this case, we have $z=0, w=2$, and $x=0$. The removed 1-agents were incident to no $(0,1)$-edges, so Equation \eqref{eq3} is satisfied. 

\item Discovery in round 4 of phase $i$ ($1 \leq i < \frac{m}{2}$):

We use the following diagram to represent the local area of the ring around the discovered successful edge. 

\begin{center}
    \begin{tikzpicture}[scale=0.75]
        \node[draw, circle, minimum size=15pt] (a) at (2,3) {};
        \node[draw, circle, minimum size=15pt] (b) at (2,1) {};
        \node[draw, circle, minimum size=15pt] (c) at (4,3) {};
        \node[draw, circle, minimum size=15pt, fill=black] (d) at (4,1) {\tiny \textcolor{white}{$a$}};
        \node[draw, circle, minimum size=15pt] (e) at (6,3) {};
        \node[draw, circle, minimum size=15pt, fill=black] (f) at (6,1) {\tiny \textcolor{white}{$b$}};
        \node[draw, circle, minimum size=15pt] (g) at (8,3) {};
        \node[draw, circle, minimum size=15pt] (h) at (8,1) {};
        \node (u) at (1,3) {};
        \node (v) at (1,2) {};
        \node (w) at (1,1) {};
        
        \node (x) at (9,3) {};
        \node (y) at (9,2) {};
        \node (z) at (9,1) {};
        
        \draw[dashed] (a) -- (b) (c) -- (d) (e) -- (f) (g) -- (h);
        \draw (v) -- (a) (b) -- (c) (d) -- (e) (f) -- (g) (h) -- (y);
        \draw (u) -- (a) (b) -- (d) (c) -- (e) (f) -- (h) (g) -- (x);
        \draw (v) -- (b) (a) -- (d) (c) -- (f) (e) -- (h) (g) -- (y);
        \draw[ultra thick] (w) -- (b) (a) -- (c) (d) -- node[yshift=-8pt] {1} (f) (e) -- (g) (h) -- (z);
        
        \draw[decoration={brace,raise=10pt},decorate] (a) -- node[yshift=20pt] {\scriptsize $i-1$} (c);
        \draw[decoration={brace,raise=10pt},decorate] (c) -- node[yshift=20pt] {\scriptsize $i-1$} (e);
        \draw[decoration={brace,raise=10pt},decorate] (e) -- node[yshift=20pt] {\scriptsize $i-1$} (g);
    \end{tikzpicture}
\end{center}
    
Here (and in all subsequent such diagrams), the edges represent pairs explored during phase $i$. The bolded matching was the last one played, i.e., the one that resulted in the 1-edge discovery.
The top brackets indicate the presence of $i-1$ columns of agents between these matched columns. All of the missing agents in the labeled gaps are deduced to be 0-agents by their proximity to the discovered 1-edge.
Discovered 1-agents are black, deduced 0-agents are white, and unknown agents are gray (in the later diagrams). 

The algorithm removes the neighborhood of $b$; i.e., the $i$ columns to the left of $b$ (including $a$) and the $i$ columns to the right of $b$. Notice that it does not remove the columns strictly to the left of $a$, even though they can be deduced to consist of 0-agents. Instead, the algorithm can deduce that all the edges between these agents and the agents to the right of $b$ are unsuccessful, and can use this deduction to patch the gap in the ring. This allows the algorithm to obtain an intermediary stage of a sub-factorization.

Both $a$ and $b$ had been paired with $4i$ 0-agents. The number of removed 0-agents is $z=4i$, and two 1-agents ($a$ and $b$) are removed, so $w=2$. $x=0$ because no known 1-agents is retained. Substituting these values, we see that Equation \eqref{eq3} is satisfied. Note that the same argument also handles the case where $(a,b)$ is a diagonal edge in our diagram, as can happen in round 4 in an odd cycle.

\item Discovery in round 3 of phase $i$ ($1 \leq i < \frac{m}{2}$):

There are 3 sub-cases to consider. 

\begin{enumerate}

\item In the first sub-case, one endpoint (w.l.o.g.~$b$) of the edge $(a,b)$ had already been connected to both vertices on the other side, while the other ($a$) had not. Let $c$ be the node on the other side that $a$ had not been connected to, and assume further that $c$ was paired with the 0-agent in the same column as $a$ in the previous round. Note that this subcase handles discovery in any even cycle, and all but two possible edge discoveries in an odd cycle. One such arrangement is shown below.

\begin{center}
    \begin{tikzpicture}[scale=0.75]
        \node[draw, circle, minimum size=15pt, fill=black!30] (a) at (2,3) {\tiny $c$};
        \node[draw, circle, minimum size=15pt] (b) at (2,1) {};
        \node[draw, circle, minimum size=15pt, fill=black] (c) at (4,3) {\tiny \textcolor{white}{$a$}};
        \node[draw, circle, minimum size=15pt] (d) at (4,1) {};
        \node[draw, circle, minimum size=15pt] (e) at (6,3) {};
        \node[draw, circle, minimum size=15pt, fill=black] (f) at (6,1) {\tiny \textcolor{white}{$b$}};
        \node[draw, circle, minimum size=15pt] (g) at (8,3) {};
        \node[draw, circle, minimum size=15pt] (h) at (8,1) {};
        \node (u) at (1,3) {};
        \node (v) at (1,2) {};
        \node (w) at (1,1) {};
        
        \node (x) at (9,3) {};
        \node (y) at (9,2) {};
        \node (z) at (9,1) {};
        
        \draw[dashed] (a) -- (b) (c) -- (d) (e) -- (f) (g) -- (h);
        \draw (v) -- (a) (b) -- (c) (d) -- (e) (f) -- (g) (h) -- (y);
        \draw (u) -- (a) (b) -- (d) (c) -- (e) (f) -- (h) (g) -- (x);
        \draw[ultra thick] (v) -- (b) (a) -- (d) (c) -- (f) (e) -- (h) (g) -- (y);
        \draw node at (5,2.5) {1};
        
        \draw[decoration={brace,raise=10pt},decorate] (a) -- node[yshift=20pt] {\scriptsize $i-1$} (c);
        \draw[decoration={brace,raise=10pt},decorate] (c) -- node[yshift=20pt] {\scriptsize $i-1$} (e);
        \draw[decoration={brace,raise=10pt},decorate] (e) -- node[yshift=20pt] {\scriptsize $i-1$} (g);
    \end{tikzpicture}
\end{center}

The algorithm plays the final matching as prescribed by \ringfac. The edge between $a$ and $c$ will identify the type of $c$. If $c$ is a 0-agent, we can appeal to Case 2. Thus, we are left to handle the case in which $c$ is a 1-agent. In that case, let $d$ and $e$ be the agents in the column $i+1$ columns to the right of $b$ in the original ring. The algorithm removes the neighborhood of $b$. In the remaining subgraph, the algorithm can deduce all unsuccessful edges needed to obtain a sub-factorization, except for the edges $(c,d)$ and $(c,e)$. Thus, the algorithm needs to do more work in the next round in order to ``catch up'' to the sub-factorization. 
In the next round, the algorithm uses $a$ and $b$ to explore $d$ and $e$ (pairing up their prescribed neighbors), while pairing up the remaining agents according to the sub-factorization as shown below. In this figure, notice that the top and bottom parts are ``interleaved'' --- for example, the nodes $d$ and $e$ are $i$ columns to the right of $c$, and are the immediate left neighbors of the gray nodes shown above them and slightly to the right.

\begin{center}
    \begin{tikzpicture}[scale=0.75]
        \node[draw, circle, minimum size=15pt, fill=black!30] (a) at (2,6) {};
        \node[draw, circle, minimum size=15pt, fill=black!30] (b) at (2,4) {\tiny $h$};
        \node[draw, circle, minimum size=15pt, fill=black] (c) at (5,6) {\color{white} \tiny $c$};
        \node[draw, circle, minimum size=15pt] (d) at (5,4) {};
        \node[draw, circle, minimum size=15pt, fill=black!30] (e) at (8,6) {};
        \node[draw, circle, minimum size=15pt, fill=black!30] (f) at (8,4) {};
        \node[draw, circle, minimum size=15pt, fill=black!30] (g) at (11,6) {};
        \node[draw, circle, minimum size=15pt, fill=black!30] (h) at (11,4) {};
        \node (v) at (0.5,5) {};
        \node (y) at (12.5,5) {};
        
        \node[draw, circle, minimum size=15pt] (c2) at (4,3) {};
        \node[draw, circle, minimum size=15pt] (d2) at (4,1) {\tiny $f$};
        \node[draw, ultra thick, circle, minimum size=15pt, fill=black!30] (e2) at (7,3) {\tiny $d$};
        \node[draw, ultra thick, circle, minimum size=15pt, fill=black!30] (f2) at (7,1) {\tiny $e$};
        \node[draw, circle, minimum size=15pt, fill=black!30] (g2) at (10,3) {\tiny $g$};
        \node[draw, circle, minimum size=15pt, fill=black!30] (h2) at (10,1) {};
        \node (v2) at (2.5,2) {};
        \node (y2) at (11.5,2) {};
        
        \draw[dashed] (a) -- (b) (c) -- (d) (e) -- (f) (e2) -- (f2) (g) -- (h) (g2) -- (h2);
        \draw[ultra thick] (v) -- (a) (v2) -- (c2) (b) -- (c) (d) -- (e) (f) -- (g) (h) -- (y) (h2) -- (y2) (d2) -- (g2);
        
        \draw[decoration={brace,raise=10pt},decorate] (a) -- node[yshift=20pt] {\scriptsize $i$} (c);
        \draw[decoration={brace,raise=10pt},decorate] (c) -- node[yshift=20pt] {\scriptsize $i$} (e);
        \draw[decoration={brace,raise=10pt},decorate] (e) -- node[yshift=20pt] {\scriptsize $i$} (g);
        \draw[decoration={brace,mirror,raise=10pt},decorate] (d2) -- node[yshift=-20pt] {\scriptsize $i$} (f2);
        \draw[decoration={brace,mirror,raise=10pt},decorate] (f2) -- node[yshift=-20pt] {\scriptsize $i$} (h2);
    \end{tikzpicture}
\end{center}

If both $d$ and $e$ are 0-agents, then the algorithm can deduce that $(c,d)$, $(c,e)$, $(f,d)$, and $(e,g)$ are unsuccessful edges, and has caught up to an intermediary stage of the ring factorization (forgetting about the explored $(e,f)$ edge). In total, $a$ was paired with $4i$ 0-agents (all $i+2$ neighbors except for $b$ and $c$) and $b$ was paired with $4i+1$ 0-agents (all $4i+2$ neighbors except $a$), while $z=4i$, $w=2$, and $x=1$, so Inequality \eqref{eq3} is satisfied.

If one of $d,e$ is a 1-agent (w.l.o.g.~$d$; they cannot both be 1-agents), then all of the $i$ columns to the right of $d, e$ consist of 0-agents. There are two further cases to consider.
\begin{itemize}
\item If $h$ is a 0-agent, then the algorithm can further remove the $2i+1$ columns surrounding $c$. Each missing edge in the sub-factorization has one endpoint that is known to be a 0-agent, so we can appeal to the inductive hypothesis. Overall, $c$ and $d$ were each paired with $4i+1$ 0-agents, and $a$ and $b$ were paired with a total of $8i$ 0-agents. In addition, $z = 8i$, $w=4$, and $x=0$, so Inequality \eqref{eq3} is satisfied.

\item If $h$ is a 1-agent, then the algorithm can remove $i-1$ columns to the left of $h$, as well as the $i+2$ columns between (and including) the columns of $h$ and $c$. Note that the missing edges in the sub-factorization each have an endpoint that is a deduced 0-agent, so they can be added to obtain an intermediary round of a smaller ring factorization. In all, $c$ was paired with $4i$ 0-agents, $h$ was paired with $(4i+1)$ 0-agents, and $a$ and $b$ were paired with a total of $8i$ 0-agents. In addition, $z=8i$, $w=4$, and $x=0$, so Inequality \eqref{eq3} is satisfied.
\end{itemize}


\item 
    In the second sub-case, as in case (a), $b$ had already been connected to both vertices on the other side, while $a$ had not. Different from case (a), we assume that $c$ was paired with an agent to its \emph{left} in the previous round. This case cannot occur in an even cycle, and can occur in exactly one place in an odd cycle. The diagram for this case is shown below.

\begin{center}
    \begin{tikzpicture}[scale=0.75]
        \node[draw, circle, minimum size=15pt, fill=black!30] (a) at (2,3) {\tiny $d$};
        \node[draw, circle, minimum size=15pt, fill=black!30] (b) at (2,1) {};
        \node[draw, circle, minimum size=15pt, fill=black!30] (c) at (4,3) {\tiny $c$};
        \node[draw, circle, minimum size=15pt] (d) at (4,1) {};
        \node[draw, circle, minimum size=15pt, fill=black] (e) at (6,3) {\tiny \textcolor{white}{$a$}};
        \node[draw, circle, minimum size=15pt] (f) at (6,1) {};
        \node[draw, circle, minimum size=15pt] (g) at (8,3) {};
        \node[draw, circle, minimum size=15pt, fill=black] (h) at (8,1) {\tiny \textcolor{white}{$b$}};
        \node (u) at (1,3) {};
        \node (v) at (1,2) {};
        \node (w) at (1,1) {};
        
        \node (x) at (9,3) {};
        \node (y) at (9,2) {};
        \node (z) at (9,1) {};
        
        \draw[dashed] (a) -- (b) (c) -- (d) (e) -- (f) (g) -- (h);
        \draw (v) -- (a) (b) -- (c) (d) -- (e) (f) -- (g) (h) -- (y);
        \draw (u) -- (a) (b) -- (d) (c) -- (f) (e) -- (g) (h) -- (z);
        \draw[ultra thick] (v) -- (b) (a) -- (c) (d) -- (f) (e) -- (h) (g) -- (y);
        \draw node at (7,2.5) {1};
        
        \draw[decoration={brace,raise=10pt},decorate] (a) -- node[yshift=20pt] {\scriptsize $i-1$} (c);
        \draw[decoration={brace,raise=10pt},decorate] (c) -- node[yshift=20pt] {\scriptsize $i-1$} (e);
        \draw[decoration={brace,raise=10pt},decorate] (e) -- node[yshift=20pt] {\scriptsize $i-1$} (g);
    \end{tikzpicture}
\end{center}

If the edge $(c,d)$ is unsuccessful, we can appeal to Case 3(a). Otherwise, the edge $(c,d)$ is successful, so $d$ is identified as a 1-agent. The algorithm then removes the neighborhoods of $c$ and $b$. The $i$ columns of 0-agents to the left of $d$ allow the algorithm to deduce all necessary 0-edges to obtain an intermediary state of the sub-factorization. Each of the $w=4$ removed 1-agents was paired with $4i-1$ 0-agents, $z=8i-2$, and $x=0$, so Inequality \eqref{eq3} is satisfied.

\item In the third sub-case, neither $a$ nor $b$ had connected to both vertices on the other side. This case cannot occur in an even cycle, and can occur in exactly one place in an odd cycle. The diagram for this case is shown below. 

\begin{center}
    \begin{tikzpicture}[scale=0.75]
        \node[draw, circle, minimum size=15pt, fill=black!30] (a) at (2,3) {\tiny $c$};
        \node[draw, circle, minimum size=15pt] (b) at (2,1) {};
        \node[draw, circle, minimum size=15pt] (c) at (4,3) {};
        \node[draw, circle, minimum size=15pt, fill=black] (d) at (4,1) {\tiny \textcolor{white}{$a$}};
        \node[draw, circle, minimum size=15pt] (e) at (6,3) {};
        \node[draw, circle, minimum size=15pt, fill=black] (f) at (6,1) {\tiny \textcolor{white}{$b$}};
        \node[draw, circle, minimum size=15pt] (g) at (8,3) {};
        \node[draw, circle, minimum size=15pt, fill=black!30] (h) at (8,1) {\tiny $d$};
        \node (u) at (1,3) {};
        \node (v) at (1,2) {};
        \node (w) at (1,1) {};
        
        \node (x) at (9,3) {};
        \node (y) at (9,2) {};
        \node (z) at (9,1) {};
        
        \draw[dashed] (a) -- (b) (c) -- (d) (e) -- (f) (g) -- (h);
        \draw (v) -- (a) (b) -- (c) (d) -- (e) (f) -- (g) (h) -- (y);
        \draw (u) -- (a) (b) -- (d) (c) -- (f) (e) -- (g) (h) -- (z);
        \draw[ultra thick] (v) -- (b) (a) -- (c) (d) -- node[yshift=-10pt] {1} (f) (e) -- (h) (g) -- (y);
        
        \draw[decoration={brace,raise=10pt},decorate] (a) -- node[yshift=20pt] {\scriptsize $i-1$} (c);
        \draw[decoration={brace,raise=10pt},decorate] (c) -- node[yshift=20pt] {\scriptsize $i-1$} (e);
        \draw[decoration={brace,raise=10pt},decorate] (e) -- node[yshift=20pt] {\scriptsize $i-1$} (g);
    \end{tikzpicture}
\end{center}

In the last round, the algorithm plays the matching prescribed by \ringfac. If both $c$ and $d$ are 0-agents, we can appeal to case 2 of the proof. If exactly one of $c,d$ is a 1-agent, we can appeal to Case 3(a). If both $c$ and $d$ are 1-agents, the algorithm removes the neighborhoods of $a$ and $d$, and patches the ring as in Case 2. $a$ and $b$ were each paired with $4i-1$ 1-agents, and $c$ and $d$ were paired with $4i$ 1-agents. In addition, $z=8i-2$, $w=4$, and $x=0$, so Inequality \eqref{eq3} is satisfied. 


\end{enumerate}

\item Discovery in round 2 of phase $i$ ($1 \leq i < \frac{m}{2}$):

There are two sub-cases to consider. 

\begin{enumerate}

\item In the first sub-case, exactly one endpoint of the discovered successful edge (w.l.o.g.~$b$) was connected to a vertex on the other side. This subcase handles discovery in any even cycle, and all but one possible edge discovery in an odd cycle. One such arrangement is shown below.

\begin{center}
    \begin{tikzpicture}[scale=0.75]
        \node[draw, circle, minimum size=15pt, fill=black!30] (a) at (2,3) {};
        \node[draw, circle, minimum size=15pt, fill=black!30] (b) at (2,1) {};
        \node[draw, circle, minimum size=15pt] (c) at (4,3) {};
        \node[draw, circle, minimum size=15pt, fill=black] (d) at (4,1) {\tiny \textcolor{white}{$a$}};
        \node[draw, circle, minimum size=15pt] (e) at (6,3) {};
        \node[draw, circle, minimum size=15pt, fill=black] (f) at (6,1) {\tiny \textcolor{white}{$b$}};
        \node[draw, circle, minimum size=15pt] (g) at (8,3) {};
        \node[draw, circle, minimum size=15pt, fill=black!30] (h) at (8,1) {};
        \node (u) at (1,3) {};
        \node (v) at (1,2) {};
        \node (w) at (1,1) {};
        
        \node (x) at (9,3) {};
        \node (y) at (9,2) {};
        \node (z) at (9,1) {};
        
        \draw[dashed] (a) -- (b) (c) -- (d) (e) -- (f) (g) -- (h);
        \draw (v) -- (a) (b) -- (c) (d) -- (e) (f) -- (g) (h) -- (y);
        \draw[ultra thick] (w) -- (b) (a) -- (c) (d) -- node[yshift=-8pt] {1} (f) (e) -- (g) (h) -- (z);
        
        \draw[decoration={brace,raise=10pt},decorate] (a) -- node[yshift=20pt] {\scriptsize $i-1$} (c);
        \draw[decoration={brace,raise=10pt},decorate] (c) -- node[yshift=20pt] {\scriptsize $i-1$} (e);
        \draw[decoration={brace,raise=10pt},decorate] (e) -- node[yshift=20pt] {\scriptsize $i-1$} (g);
    \end{tikzpicture}
\end{center}

The algorithm removes $b$'s column, along with the $i$ columns to its left and the $i-1$ columns to its right. Note that the $i-1$ columns of known 0-agents to the left of $a$ and the known 0-agent with whom $b$ was paired in the first round of this phase allow the algorithm to deduce all necessary 0-edges to obtain an intermediary stage of the sub-factorization. $a$ and $b$ were each paired with $4i-2$ 0-agents, $z=4i-2$, $w=2$, and $x=0$, so Inequality \eqref{eq3} is satisfied.


\item In this sub-case, both $a$ and $b$ had been connected to an agent on their other side. This case cannot occur in an even cycle, and can occur in exactly one place in an odd cycle. The diagram for this case is shown below.

\begin{center}
    \begin{tikzpicture}[scale=0.75]
        \node[draw, circle, minimum size=15pt, fill=black!30] (a) at (2,3) {};
        \node[draw, circle, minimum size=15pt] (b) at (2,1) {};
        \node[draw, circle, minimum size=15pt, fill=black] (c) at (4,3) {\tiny \textcolor{white}{$a$}};
        \node[draw, circle, minimum size=15pt] (d) at (4,1) {};
        \node[draw, circle, minimum size=15pt] (e) at (6,3) {};
        \node[draw, circle, minimum size=15pt, fill=black] (f) at (6,1) {\tiny \textcolor{white}{$b$}};
        \node[draw, circle, minimum size=15pt] (g) at (8,3) {};
        \node[draw, circle, minimum size=15pt, fill=black!30] (h) at (8,1) {};
        \node (u) at (1,3) {};
        \node (v) at (1,2) {};
        \node (w) at (1,1) {};
        
        \node (x) at (9,3) {};
        \node (y) at (9,2) {};
        \node (z) at (9,1) {};
        
        \draw[dashed] (a) -- (b) (c) -- (d) (e) -- (f) (g) -- (h);
        \draw (v) -- (a) (b) -- (c) (d) -- (e) (f) -- (g) (h) -- (y);
        \draw[ultra thick] (u) -- (a) (b) -- (d) (c) -- node[yshift=10pt] {1} (f) (e) -- (g) (h) -- (z);
        
        \draw[decoration={brace,raise=10pt},decorate] (a) -- node[yshift=20pt] {\scriptsize $i-1$} (c);
        \draw[decoration={brace,raise=10pt},decorate] (c) -- node[yshift=20pt] {\scriptsize $i-1$} (e);
        \draw[decoration={brace,raise=10pt},decorate] (e) -- node[yshift=20pt] {\scriptsize $i-1$} (g);
    \end{tikzpicture}
\end{center}

Again, the algorithm removes $b$'s column, along with the $i$ columns to its left and the $i-1$ columns to its right. The $i-1$ columns of 0-agents to the left of $a$, along with the known 0-agents with whom $a$ and $b$ were paired in the first round of this phase, allow the algorithm to deduce all necessary 0-edges to obtain an intermediary stage of the sub-factorization. $a$ and $b$ were each paired with $4i-2$ 0-agents, $z=4i-2$, $w=2$, and $x=0$, so Inequality \eqref{eq3} is satisfied.
\end{enumerate}

\item Discovery in round 1 of phase $i$ ($1 \leq i < \frac{m}{2}$):
Our diagram is as follows; note that in this case, we do not use $a$ and $b$ to denote the discovered successful edge:

\begin{center}
    \begin{tikzpicture}[scale=0.75]
        \node[draw, circle, minimum size=15pt, fill=black!30] (a) at (2,3) {\tiny $a$};
        \node[draw, circle, minimum size=15pt, fill=black!30] (b) at (2,1) {\tiny $b$};
        \node[draw, circle, minimum size=15pt] (c) at (4,3) {};
        \node[draw, circle, minimum size=15pt, fill=black] (d) at (4,1) {};
        \node[draw, circle, minimum size=15pt, fill=black] (e) at (6,3) {};
        \node[draw, circle, minimum size=15pt] (f) at (6,1) {};
        \node[draw, circle, minimum size=15pt, fill=black!30] (g) at (8,3) {\tiny $c$};
        \node[draw, circle, minimum size=15pt, fill=black!30] (h) at (8,1) {\tiny $d$};
        \node (u) at (1,3) {};
        \node (v) at (1,2) {};
        \node (w) at (1,1) {};
        
        \node (x) at (9,3) {};
        \node (y) at (9,2) {};
        \node (z) at (9,1) {};
        
        \draw[dashed] (a) -- (b) (c) -- (d) (e) -- (f) (g) -- (h);
        \draw[ultra thick] (v) -- (a) (b) -- (c) (d) -- node[yshift=10pt] {1} (e) (f) -- (g) (h) -- (y);
        
        \draw[decoration={brace,raise=10pt},decorate] (a) -- node[yshift=20pt] {\scriptsize $i-1$} (c);
        \draw[decoration={brace,raise=10pt},decorate] (c) -- node[yshift=20pt] {\scriptsize $i-1$} (e);
        \draw[decoration={brace,raise=10pt},decorate] (e) -- node[yshift=20pt] {\scriptsize $i-1$} (g);
    \end{tikzpicture}
\end{center}

The algorithm begins by removing the column of the right 1-agent, as well as the $i$ columns to its left and the $i-1$ columns to its right. At this point, the algorithm is almost left with an intermediary stage of a sub-factorization; however, it cannot deduce whether the edge $(b,c)$ is successful (which is necessary to recover a sub-factorization). To ``catch up'' to the sub-factorization, the algorithm does some exploration with the removed 1-agents. In the second round, \ringfac prescribes a pairing of $(a,c)$ or $(b,d)$.\footnote{In an even cycle, all round-2 edges are within the same ring. In an odd cycle, there will be one ``diagonal'' edge connecting nodes from different rings played in round 2. However, since the exploration graph is rotationally symmetric after round 1, the algorithm is free to choose the location of this diagonal edge; in particular, it places the edge between columns other than $b$'s and $c$'s.} The algorithm uses the removed 1-agents to explore both endpoints of this edge, while playing the rest of the matching according to the sub-factorization. One possibility is shown below; by the footnote, it suffices w.l.o.g.~to focus on this case. 

\begin{center}
    \begin{tikzpicture}[scale=0.75]
        \node[draw, circle, minimum size=15pt, fill=black!30] (a) at (2,3) {};
        \node[draw, circle, minimum size=15pt, fill=black!30] (b) at (2,1) {\tiny $e$};
        \node[draw, circle, minimum size=15pt, fill=black!30] (c) at (4,3) {\tiny $a$};
        \node[draw, circle, ultra thick, minimum size=15pt, fill=black!30] (d) at (4,1) {\tiny $b$};
        \node[draw, circle, minimum size=15pt, fill=black!30] (e) at (6,3) {\tiny $c$};
        \node[draw, circle, ultra thick, minimum size=15pt, fill=black!30] (f) at (6,1) {\tiny $d$};
        \node[draw, circle, minimum size=15pt, fill=black!30] (g) at (8,3) {\tiny $f$};
        \node[draw, circle, minimum size=15pt, fill=black!30] (h) at (8,1) {};
        \node (u) at (1,3) {};
        \node (v) at (1,2) {};
        \node (w) at (1,1) {};
        
        \node (x) at (9,3) {};
        \node (y) at (9,2) {};
        \node (z) at (9,1) {};
        
        \draw[dashed] (a) -- (b) (c) -- (d) (e) -- (f) (g) -- (h);
        \draw (v) -- (a) (b) -- (c) (f) -- (g) (h) -- (y);
        \draw[ultra thick] (w) -- (b) (a) -- (c) (e) -- (g) (h) -- (z);
        
        \draw[decoration={brace,raise=10pt},decorate] (a) -- node[yshift=20pt] {\scriptsize $i-1$} (c);
        \draw[decoration={brace,raise=10pt},decorate] (c) -- node[yshift=20pt] {\scriptsize $i-1$} (e);
        \draw[decoration={brace,raise=10pt},decorate] (e) -- node[yshift=20pt] {\scriptsize $i-1$} (g);
    \end{tikzpicture}
\end{center}

If $b$ is a 0-agent, then the algorithm can deduce the missing edge $(b,c)$, and is back at an intermediary stage of the sub-factorization. The removed 1-agents were each matched to at most $4i-2$ 0-agents, $z=4i-2$, $w=2$, and $x \geq 0$, so Inequality \eqref{eq3} is satisfied.

Thus, it remains to handle cases in which $b$ is a 1-agent.
If the edge $(c,f)$ is successful, then the algorithm removes $b$'s column, along with the $i$ columns to its right and $i-2$ columns to its left. The originally removed 1-agents were paired with a total of $8i-5$ 0-agents, $b$ was paired with $4i-2$ 0-agents, and $c$ with $4i-3$ 0-agents. In addition, $z=8i-6$, $w=4$, and $x=1$, so Inequality \eqref{eq3} is satisfied. A similar procedure handles the cases in which $d$ is a 1-agent, or in which $e$'s edge in round 2 is successful.
If none of these cases arises, the algorithm proceeds with the next round, as shown in the following diagram.

\begin{center}
    \begin{tikzpicture}[scale=0.75]
        \node[draw, circle, minimum size=15pt, fill=black!30] (a) at (2,3) {};
        \node[draw, circle, minimum size=15pt, fill=black!30] (b) at (2,1) {\tiny $e$};
        \node[draw, circle, ultra thick, minimum size=15pt] (c) at (4,3) {\tiny $a$};
        \node[draw, circle, minimum size=15pt, fill=black] (d) at (4,1) {\tiny \textcolor{white}{$b$}};
        \node[draw, circle, ultra thick, minimum size=15pt, fill=black!30] (e) at (6,3) {\tiny $c$};
        \node[draw, circle, minimum size=15pt] (f) at (6,1) {\tiny $d$};
        \node[draw, circle, minimum size=15pt, fill=black!30] (g) at (8,3) {\tiny $f$};
        \node[draw, circle, minimum size=15pt, fill=black!30] (h) at (8,1) {};
        \node (u) at (1,3) {};
        \node (v) at (1,2) {};
        \node (w) at (1,1) {};
        
        \node (x) at (9,3) {};
        \node (y) at (9,2) {};
        \node (z) at (9,1) {};
        
        \draw[dashed] (a) -- (b) (c) -- (d) (e) -- (f) (g) -- (h);
        \draw (v) -- (a) (b) -- (c) (f) -- (g) (h) -- (y);
        \draw (w) -- (b) (a) -- (c) (d) -- (f) (e) -- (g) (h) -- (z);
        \draw[ultra thick] (u) -- (a) (b) -- (d) (f) -- (h) (g) -- (x);
        
        \draw[decoration={brace,raise=10pt},decorate] (a) -- node[yshift=20pt] {\scriptsize $i-1$} (c);
        \draw[decoration={brace,raise=10pt},decorate] (c) -- node[yshift=20pt] {\scriptsize $i-1$} (e);
        \draw[decoration={brace,raise=10pt},decorate] (e) -- node[yshift=20pt] {\scriptsize $i-1$} (g);
    \end{tikzpicture}
\end{center}

If $c$ is revealed to be a 0-agent, then the algorithm can deduce the missing edge $(b,c)$ to be unsuccessful, and is back at an intermediary stage of the sub-factorization. In this case, the two originally removed 1-agents had been matched to a total of $8i-3$ 0-agents, $z=4i-2$, $w=2$, and $x = 1$, so Inequality \eqref{eq3} is satisfied. We are left to handle cases where $c$ is a 1-agent, implying that the $i-1$ columns to its right only contain 0-agents. Note that because we are in the case where the edge $(c,f)$ was unsuccessful, this implies that $f$ was a 0-agent.

If $e$ is a 0-agent (which is revealed by the edge with $b$), the algorithm removes $b$'s column, along with the $i-1$ columns to its left and the $i$ columns to its right. The known 0-agents to the right of $c$, along with $e$ being a 0-agent, allow the algorithm to infer all missing edges to be unsuccessful. The originally removed 1-agents had been matched to a total of $8i-4$ 0-agents, and $b$ and $c$ had each been matched to $4i-2$ 0-agents. In addition, $z=8i-4$, $w=4$, and $x=0$, so Inequality \eqref{eq3} is satisfied.

Otherwise, $e$ is a 1-agent. In this case, the algorithm removes $e$'s column, along with the $i-1$ columns to its left and the $i$ columns to its right. Because the $i-1$ columns to the right of $c$, as well as the agent $f$, are known to be 0-agents, this allows the algorithm to infer all missing edges to catch up with a sub-factorization. The originally removed 1-agents had been matched to a total of $8i-4$ 0-agents, $b$ was matched to $4i-2$ and $e$ to $4i-1$. In addition, $z=8i-4$, $w=4$, and $x=1$ (because $c$ had been matched with $b$), so Inequality \eqref{eq3} is again satisfied.

\item Discovery is made during phase $\frac{m}{2}$ ($m$ even):

In this case, $\K=2$. Since no 1-agent is ever paired to the same 0-agent multiple times, the total regret is at most $\frac{1}{2} \cdot \K \cdot (\N-\K) = \N-\K = \N - \K + \left\lfloor \frac{\K}{4} \right\rfloor$. 
\end{enumerate}

We have accounted for all possible scenarios where the first 1-edge is discovered. Therefore, we have completed the proof of Lemma~\ref{lem:removal-bound}.
\end{extraproof}

%% file: sa6_104.tex
\section[A Weakest Link Instance with Regret Exceeding n-k]{A Weakest Link Instance with Regret Exceeding $\N - \K$} \label{app:104}

We consider the instance of the weakest link (\textsf{AND}) model with $\N=10$ agents, $\K=4$ of whom have high skill. We argue that any algorithm on this instance can be made to incur regret 7. Note that this is strictly more regret than the $\N-\K = 6$ guaranteed by \Cref{thm:andlb}. It also exactly matches the upper regret bound from \Cref{thm:andub}. Hence, the \ringalg~Algorithm is optimal for this instance. 

Regardless of the actions of the algorithm in rounds 1--2, the adversary will reveal no $(1,1)$-teams. This is always possible since the exploration graph after round 2 is bipartite, so it includes an independent set of size 5. Next, we describe the actions of the adversary in round 3. To do this, we consider the possible exploration graphs $G(V,E_3)$. By a computer search, we determined that there are 102 non-isomorphic exploration graphs.

97 of these graphs have two (not necessarily disjoint) independent sets $I_1, I_2$ with $|I_1| = |I_2| = 4$ and such that $|I_1 \cup I_2|$ is odd. The adversary can use $I_1$ and $I_2$ to force the algorithm to incur regret at least 7. First, the existence of $I_1$ and $I_2$ allows the adversary to again reveal no $(1,1)$-pairs in round 3. At the end of round 3, the algorithm has accrued 6 units of regret. Since $I_1 \cup I_2$ has odd cardinality, any matching that the algorithm chooses must include an edge connecting some agent in $I_1 \cup I_2$ to an agent $w \notin I_1 \cup I_2$. However, this gives the adversary a labeling that includes a $(0,1)$-team in round 4, by making either all of $I_1$ or all of $I_2$ have type 1, depending on which of the two has the edge to $w$. This $(0,1)$-edge adds at least one additional unit of regret. We draw these 97 graphs below, using red and blue dots to depict $I_1$ and $I_2$. 

In the remaining 5 graphs (the last 5 graphs shown below), either (i) there is no independent set of size 4 or (ii) there is a matching such that for every independent set of size 4, each edge in the matching either connects two agents in the independent set or two agents not in the independent set. In either case, the adversary must reveal a $(1,1)$-team in round 3, either to ensure a consistent type assignment or to force total regret 7.

For each of these graphs, we programmatically verified the following:
\begin{enumerate}
    \item Given any three matchings whose union is the depicted graph, each of the three matchings includes one of the blue edges (or the image of one of the bolded blue edges under an automorphism of the graph). 
    \item By revealing the blue edge contained in the round-3 matching as a $(1,1)$-edge, the adversary can always either (i) ensure regret 2 in round 4 (so the regret in round 3 is one, and in round 4, it is two) or (ii) ensure regret 1 in round 4 and leave two independent 2-sets $I_1, I_2$ in the unresolved subgraph such that $|I_1 \cup I_2| = 3$ (so the regret in each of rounds 3,4, and 5 is one).
\end{enumerate}

Thus, the total regret in each of the cases is at least 7.

\begin{figure}[H]
    \begin{minipage}{0.16\textwidth}

    \end{minipage}
    \hspace{25pt}
    \begin{minipage}{0.16\textwidth}
    \end{minipage}
    \hspace{25pt}
    \begin{minipage}{0.16\textwidth}
    \end{minipage}
    \hspace{25pt}
    \begin{minipage}{0.16\textwidth}
    \end{minipage}
\end{figure}

%% file: paper.bbl
\begin{thebibliography}{10}

\bibitem{Aigner95_Turan}
Martin Aigner.
\newblock Tur\'{a}n's graph theorem.
\newblock {\em The American Mathematical Monthly}, 102(9):808--816, 1995.

\bibitem{babaioff2006combinatorial}
Moshe Babaioff, Michal Feldman, and Noam Nisan.
\newblock Combinatorial agency.
\newblock In {\em Proc. 7th ACM Conf. on Electronic Commerce}, page 18–28,
  2006.

\bibitem{carlier2010matching}
Guillaume Carlier and Ivar Ekeland.
\newblock Matching for teams.
\newblock {\em Economic Theory}, 42:397--418, 02 2010.

\bibitem{cesa2012combinatorial}
Nicol\`{o} Cesa-Bianchi and G\'{a}bor Lugosi.
\newblock Combinatorial bandits.
\newblock {\em Journal of Computer and System Sciences}, 78(5):1404--1422,
  2012.

\bibitem{Chen2013CombinatorialMB}
Wei Chen, Yajun Wang, and Yang Yuan.
\newblock Combinatorial multi-armed bandit: General framework and applications.
\newblock In {\em ICML}, 2013.

\bibitem{combes2015combinatorial}
Richard Combes, M.~Sadegh Talebi, Alexandre Proutiere, and Marc Lelarge.
\newblock Combinatorial bandits revisited.
\newblock In {\em Proc. 29th Advances in Neural Information Processing
  Systems}, page 2116–2124, 2015.

\bibitem{devanur2012online}
Nikhil~R Devanur and Kamal Jain.
\newblock Online matching with concave returns.
\newblock In {\em Proceedings of the forty-fourth annual ACM symposium on
  Theory of computing}, pages 137--144, 2012.

\bibitem{gai2012combinatorial}
Yi~Gai, Bhaskar Krishnamachari, and Rahul Jain.
\newblock Combinatorial network optimization with unknown variables:
  Multi-armed bandits with linear rewards and individual observations.
\newblock {\em IEEE/ACM Transactions on Networking}, 20(5):1466--1478, 2012.

\bibitem{gopalan2014thompson}
Aditya Gopalan, Shie Mannor, and Yishay Mansour.
\newblock Thompson sampling for complex online problems.
\newblock In {\em International conference on machine learning}, pages
  100--108. PMLR, 2014.

\bibitem{han2021adversarial}
Yanjun Han, Yining Wang, and Xi~Chen.
\newblock Adversarial combinatorial bandits with general non-linear reward
  functions.
\newblock In {\em International Conference on Machine Learning}, pages
  4030--4039, 2021.

\bibitem{hssaine2018information}
Chamsi Hssaine and Siddhartha Banerjee.
\newblock Information signal design for incentivizing team formation.
\newblock In {\em Proc. 14th Conference on Web and Internet Economics (WINE)},
  2018.

\bibitem{Johari18_Team}
Ramesh Johari, Vijay Kamble, Anilesh~K. Krishnaswamy, and Hannah Li.
\newblock Exploration vs.~exploitation in team formation.
\newblock In {\em Proc. 14th Conference on Web and Internet Economics (WINE)},
  2018.

\bibitem{katariya2017stochastic}
Sumeet Katariya, Branislav Kveton, Csaba Szepesvari, Claire Vernade, and Zheng
  Wen.
\newblock Stochastic rank-1 bandits.
\newblock In {\em Artificial Intelligence and Statistics}, pages 392--401.
  PMLR, 2017.

\bibitem{kirschner2020information}
Johannes Kirschner, Tor Lattimore, and Andreas Krause.
\newblock Information directed sampling for linear partial monitoring.
\newblock In {\em Conference on Learning Theory}, pages 2328--2369. PMLR, 2020.

\bibitem{kleinberg2015team}
Jon Kleinberg and Maithra Raghu.
\newblock Team performance with test scores.
\newblock In {\em Proc. 16th ACM Conf. on Economics and Computation}, 2015.

\bibitem{kveton2015tight}
Branislav Kveton, Zheng Wen, Azin Ashkan, and Csaba Szepesv{\'{a}}ri.
\newblock Tight regret bounds for stochastic combinatorial semi-bandits.
\newblock In {\em Proc. 18th Intl. Conf. on Artificial Intelligence and
  Statistics}, 2015.

\bibitem{merlis2020tight}
Nadav Merlis and Shie Mannor.
\newblock Tight lower bounds for combinatorial multi-armed bandits.
\newblock In {\em Conference on Learning Theory}, pages 2830--2857. PMLR, 2020.

\bibitem{Rajkumar17_Partition}
Arun Rajkumar, Koyel Mukherjee, and Theja Tulabandhula.
\newblock Learning to partition using score based compatibilities.
\newblock In {\em Proc. 16th Intl. Conf. on Autonomous Agents and Multiagent
  Systems}, page 574–582, 2017.

\bibitem{russo2014learning}
Daniel Russo and Benjamin Van~Roy.
\newblock Learning to optimize via information-directed sampling.
\newblock {\em Advances in Neural Information Processing Systems}, 27, 2014.

\bibitem{sentenac2021pure}
Flore Sentenac, Jialin Yi, Clement Calauzenes, Vianney Perchet, and Milan
  Vojnovic.
\newblock Pure exploration and regret minimization in matching bandits.
\newblock In {\em ICML21}, pages 9434--9442, 2021.

\bibitem{singla2015learning}
Adish Singla, Eric Horvitz, Pushmeet Kohli, and Andreas Krause.
\newblock Learning to hire teams.
\newblock In {\em Third AAAI Conference on Human Computation and
  Crowdsourcing}, 2015.

\bibitem{Turan41}
Paul Tur{\'a}n.
\newblock Egy gr{\'a}felm{\'e}leti sz{\'e}lso{\'e}rt{\'e}kfeladatr{\'o}l (on an
  extremal problem in graph theory).
\newblock {\em Mat. Fiz. Lapok}, 48(3):436--453, 1941.

\bibitem{wang2018thompson}
Siwei Wang and Wei Chen.
\newblock Thompson sampling for combinatorial semi-bandits.
\newblock In {\em International Conference on Machine Learning}, pages
  5114--5122. PMLR, 2018.

\bibitem{zimmert2018factored}
Julian Zimmert and Yevgeny Seldin.
\newblock Factored bandits.
\newblock {\em Advances in Neural Information Processing Systems}, 31, 2018.

\end{thebibliography}
